\documentclass[12pt]{article}
\usepackage{graphicx}% Include figure files
\usepackage{dcolumn}% Align table columns on decimal point
\usepackage{bm}% bold math\usepackage{latexsym}
\usepackage{graphics}
\usepackage{subfigure}
\begin{document}
%%%%%%%%%%%%%%%%%

\begin{center}

{\bf{\large PDF in PDFs from Hadronic Tensor and LaMET}}
%Valence and Connected Sea Partons   (PDF in PDFs) in (from) \\
%\vspace{0.2cm}
%Hadronic Tensor and LaMET}}

\vspace{0.6cm}

%{\bf $\chi$QCD Collaboration}

{\bf  Keh-Fei Liu}

\end{center}

\begin{center}
\begin{flushleft}
{\it
Department of Physics and Astronomy, University of Kentucky, Lexington, KY 40506
}
\end{flushleft}
\end{center}

\begin{abstract}
We point out a problem of the phenomenological definition of the valence partons as the difference between the partons and antipartons in the context of the NNLO evolution equations. After  demonstrating  that the classification of the parton degrees of freedom (PDF) of the parton distribution functions (PDFs) are the same in the QCD path-integral formulations of the hadronic tensor and the quasi-PDF with LaMET, we resolve the problem by showing that the proper definition of the valence should be in terms of those in the connected insertions only. We also prove that the strange partons appear as the disconnected sea in the nucleon. 
\end{abstract}

%\maketitle

\section{Introduction} \label{intro}

Partonic structure of the nucleon has been discovered and extensively studied in deep inelastic 
scattering (DIS) of leptons. Further experiments in Drell-Yan process, semi-inclusive DIS (SIDIS) help to identify and clarify the flavor dependence, particularly the sea partons~\cite{Peng:2014hta}.  The first experimental evidence that the sea patrons have non-trivial flavor dependence is revealed in the experimental demonstration of the violation of Gottfried sum rule. The original 
Gottfried sum rule, $I_G \equiv \int^1_0 dx [F^p_2(x)-F^n_2(x)]/x  =1/3$, was obtained
under the assumption that $\bar u$ and $\bar d$ sea partons are the same~\cite{gottfried}. However, the NMC measurement~\cite{nmc} 
of $\int^1_0 dx [F^p_2(x)-F^n_2(x)]/x$ turns out to be $0.235 \pm 0.026$, a 4 $\sigma$ difference from the Gottfried sum rule, which implies that the $\bar u = \bar d$ assumption was invalid. Other 
flavor-dependent issues under active experimental and theoretical pursuits include the intrinsic strange and charm partons~\cite{Brodsky:1980pb}, and the 
$s(x) - \bar{s}(x)$~\cite{Davidson:2001ji,Kretzer:2003wy} and $c(x) - \bar{c}(x)$~\cite{Sufian:2020coz} asymmetries. 

In this work, we shall scrutinize the definition of the valence parton in the context of \mbox{$s(x) - \bar{s}(x)$} asymmetry and the NNLO evolution. The conventional phenomenological definition of the valence
parton is $q^-(x) \equiv q(x) - \bar{q}(x)$. Given this definition, one faces the following questions:
\begin{enumerate}
\item 
It has been suggested that the `NuTeV anomaly'~\cite{Zeller:2001hh} might be resolved if there is an
$s(x) - \bar{s}$ asymmetry~\cite{Davidson:2001ji,Kretzer:2003wy}. Even though the global analyses do not have a definite conclusion yet, when NNLO is considered for parton evolution, there is a term in the splitting function which will generate $s(x) - \bar{s}$ asymmetry. This involves a quark loop with three gluon lines attached to it. It is small, i.e.  $\mathcal{O}(\alpha_s^3)$, but non-zero nonetheless. Hence the question: although there is no net strangeness in the nucleon (i.e. $\int dx (s(x) - \bar{s}) = 0)$, why should the strange parton with  $s^-(x) = s(x) - \bar{s}(x) \neq 0$ be a part of the valence distribution.  This is contrary to the picture of the quark model, particularly its $SU(6)$ classification of the hadrons, in which the nucleon is composed of $u$ and $d$ valence quarks only, while the strange is part of the sea. 
 %Furthermore, in lattice
 %calculation, the interpolation field for the nucleon do not include strange quarks usually.

\item  
Similarly, there is a question about the NNLO evolution. One of the NNLO equations for 
$q^-(x)$~\cite{Moch:2004pa,Cafarella:2008du,Nadolsky2011} is
\begin{equation}   \label{q-}
\frac{dq_i^-}{dt}=  (P_{qq}^v - P_{q\bar{q}}^v) \otimes q_i^- + (P_{qq}^s - P_{q\bar{q}}^s) \otimes \Sigma_v, 
\end{equation}
 where $t = \ln \mu^2$. $P_{qq}^-$, $P_{qq}^s$ and $P_{q\bar{q}}^s$ are splitting functions, and
\begin{eqnarray}   \label{q-def}
\Sigma_v \equiv \sum_{i} (q_i - \bar{q}_i);  \hspace{1cm} q_i^- \equiv q_i - \bar{q}_i.
\end{eqnarray} 
When and if $q_i^-$ is interpreted as 
the valence distribution, the second term on the right of Eq.~(\ref{q-}) implies that the valence distributions of $d$ and $s$ can affect the evolution of the valence $u$ parton. This appears to be contradictory to the fact that there is no flavor-changing couplings in QCD between the valence quarks. 
\end{enumerate}

These concerns are hints that something is wrong with identifying $q^-$ as the valence distribution.
It turns out these puzzles can be resolved via the Euclidean path-integral formulation of QCD.

This manuscript is organized as follows. Sec.~\ref{HTparton} gives the path-integral formulation of the hadronic tensor which defines the parton degrees of freedom. Sec.~\ref{Feynman-x} gives the quasi-parton approach to calculating the PDF in Feynman-x directly on the lattice via large momentum effective theory (LaMET). It is shown that the parton degrees of the freedom are identical to those from the hadronic tensor. We shall present 
the resolution of the above puzzles in Sec.~\ref{resolution}. Finally, we prove in Sec.~\ref{strange} that the strange quarks only appear in the disconnected insertions in nucleon matrix elements. The summary is given in Sec.~\ref{summary}.

\section{Euclidean Path-integral Formulation of the Hadronic Tensor}  \label{HTparton}

\begin{figure}[ht]  
% \vspace*{1cm}
  \centering
% \hspace{2cm}
 \includegraphics[width=0.6\hsize]{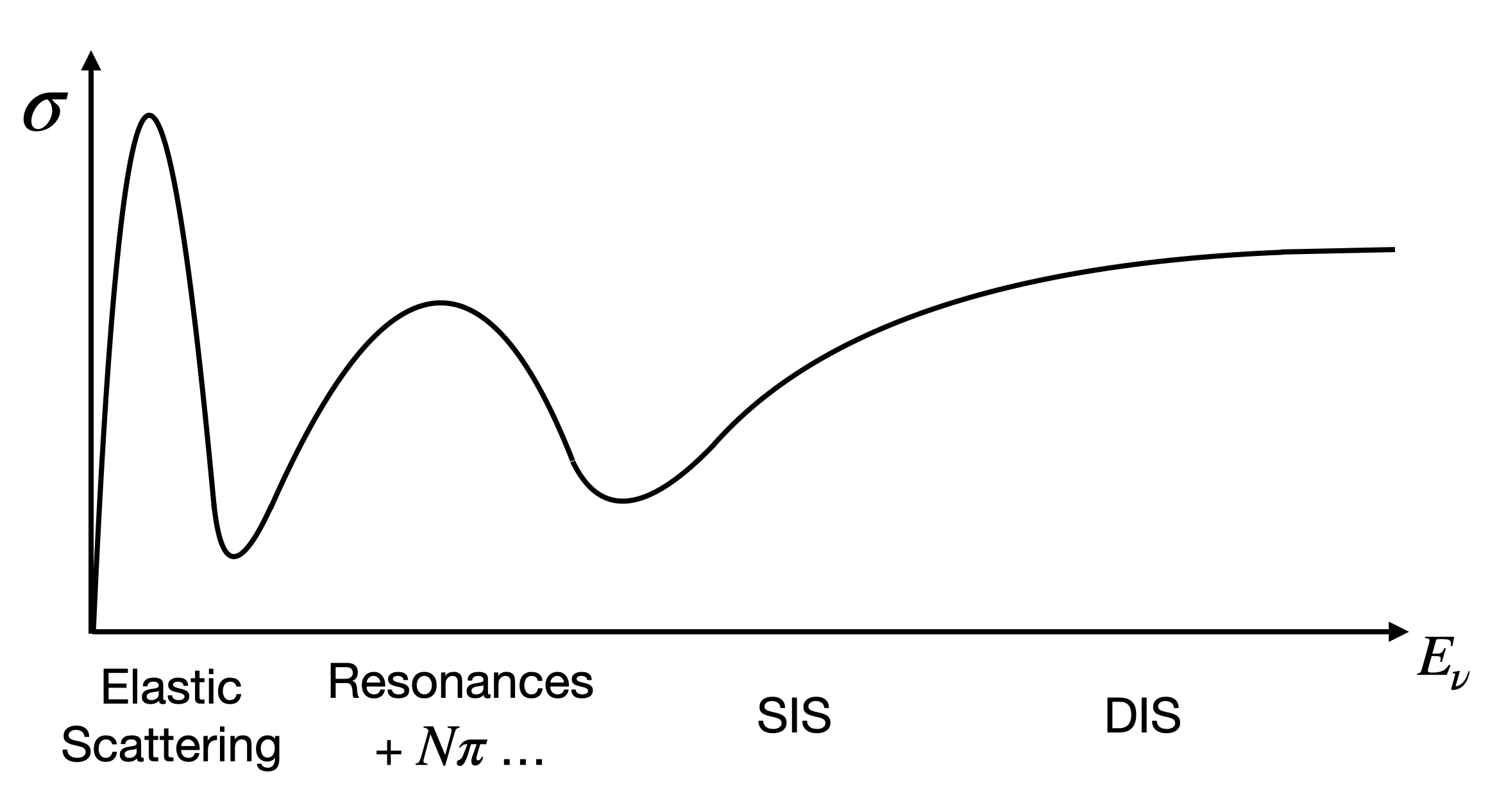}
\caption{Illustrated spectral density of the cross-section or structure function of the $\nu$-N scattering  to show the elastic, the resonance, the SIS, and the DIS regions at different energy transfer $\nu$.  }
 \label{nu-N-spectrum}
 \end{figure}

The Euclidean hadronic tensor was formulated in the path-integral formalism to identify the origin of the Gottfried sum rule violation~\cite{Liu:1993cv}. It is the current-current correlator in the nucleon and can
be obtained by the following four-point-to-two-point correlator ratio
\begin{eqnarray}  \label{wmunu_tilde}
\widetilde{W}_{\mu\nu}(\vec{q},\vec{p},\tau) &=&
 \frac{E_p }{m_N} \frac{{\rm Tr} (\Gamma_e G_{pWp}(t_0,t_1,t_2,t_f))}{{\rm Tr} (\Gamma_e G_{pp}(t_0,t_f))}
 \begin{array}{|l} \\  \\  t _f -t_2 \gg 1/\Delta E_p, \, t_1 - t_0 \gg 1/\Delta E_p \end{array} \nonumber \\
    &=& <N(\vec{p})| \int d^3x  \frac{e^{i \vec{q}\cdot \vec{x}}}{4\pi} e^{-i\vec{q}\cdot \vec{x}}
J_{\mu}(\vec{x},\tau) J_{\nu}(0,0)|N(\vec{p})>,
\end{eqnarray}
where $\tau = t_2 - t_1$ is the Euclidean time separation between the current $J_{\nu}(t_2)$ and 
$J_{\mu}(t_1)$. The current-source and sink-current separations $t_1 - t_0$ and $t_f - t_2$ are much larger than the inverse of the energy $\Delta E_p$ between the nucleon and its first excited state, so that the nucleon excited states are filtered out at large time separations. Formally, the inverse Laplace transform converts $\widetilde{W}_{\mu\nu}(\vec{q},\vec{p},\tau)$ to the Minkowski hadronic tensor
\begin{equation}  \label{wmunu}  
W_{\mu\nu}(\vec{q},\vec{p},\nu) = \frac{1}{2m_Ni} \int_{c-i \infty}^{c+i \infty} d\tau\,
e^{\nu\tau} \widetilde{W}_{\mu\nu}(\vec{q},\vec{p}, \tau),
\end{equation} 
with $c > 0$. This is basically doing the anti-Wick rotation to go back to the Minkowski space. 
In practice with the lattice calculation, it is not possible to perform the inverse Laplace transform in Eq.~(\ref{wmunu}), as there are no data on the imaginary $\tau$. Instead, one can turn this into an inverse problem and find a solution from the Laplace transform~\cite{Liu:2016djw}
  \begin{equation}  \label{Laplace}
 \widetilde{W}_{\mu\nu} (\vec{q},\vec{p},\tau) = \int d \nu\, e^{-\nu\tau}
 W_{\mu\nu}(\vec{q},\vec{p},\nu).
  \end{equation}
  This has been studied~\cite{Liu:2016djw,Liang:2017mye,Liang:2019frk,Liang:2020sqi} with the inverse algorithms such as 
Backus-Gilbert,  Maximum Entropy  and Bayesian Reconstruction methods. 
The expected spectral density of the neutrino-nucleon scattering cross section or structure
functions is illustrated in Fig.~\ref{nu-N-spectrum},
 which shows that there are several kinematic regions in the spectral density in the energy 
 transfer \mbox{$\nu$ -- the} elastic scattering, the inelastic reactions ($\pi N, \pi\pi M, \eta N$ etc.) and resonances ($\Delta$, Roper, $S_{11}$, etc.), shallow inelastic scattering (SIS), and deep inelastic scattering  (DIS) regions. To determine how large a $\nu$ is needed for DIS, we look at $W$, the total invariant mass of the hadronic state for the nucleon target at rest
\begin{equation}
W^2 = (q+p)^2 = m_p^2 - Q^2 + 2 m_p \nu
\end{equation}
 The global analyses of the high energy lepton-nucleon and Drell-Yan experiments to extract the parton distribution functions (PDFs) usually make a cut with $W^2 > 10\, {\rm GeV^2}$ to avoid the
 elastic and inelastic regions. Thus, to be qualified in the DIS region, the energy transfer  $\nu$ needs to be 
 \begin{equation}  \label{nu}
 \nu\, > 4.86\, {\rm GeV} + 0.533\,(\rm GeV^{-1})\, Q^2
\end{equation}
If we take  $Q^2 = 4\, {\rm GeV^2}$, this implies $\nu > 7$ GeV. It is shown recently in a lattice calculation that small lattice spacing (e.g. $a \le 0.04$ fm) is needed to reach such high energy 
 excitations on the lattice~\cite{Liang:2019frk}. 

\begin{figure}  
\centering
\subfigure[]
{{\includegraphics[width=0.32\hsize]{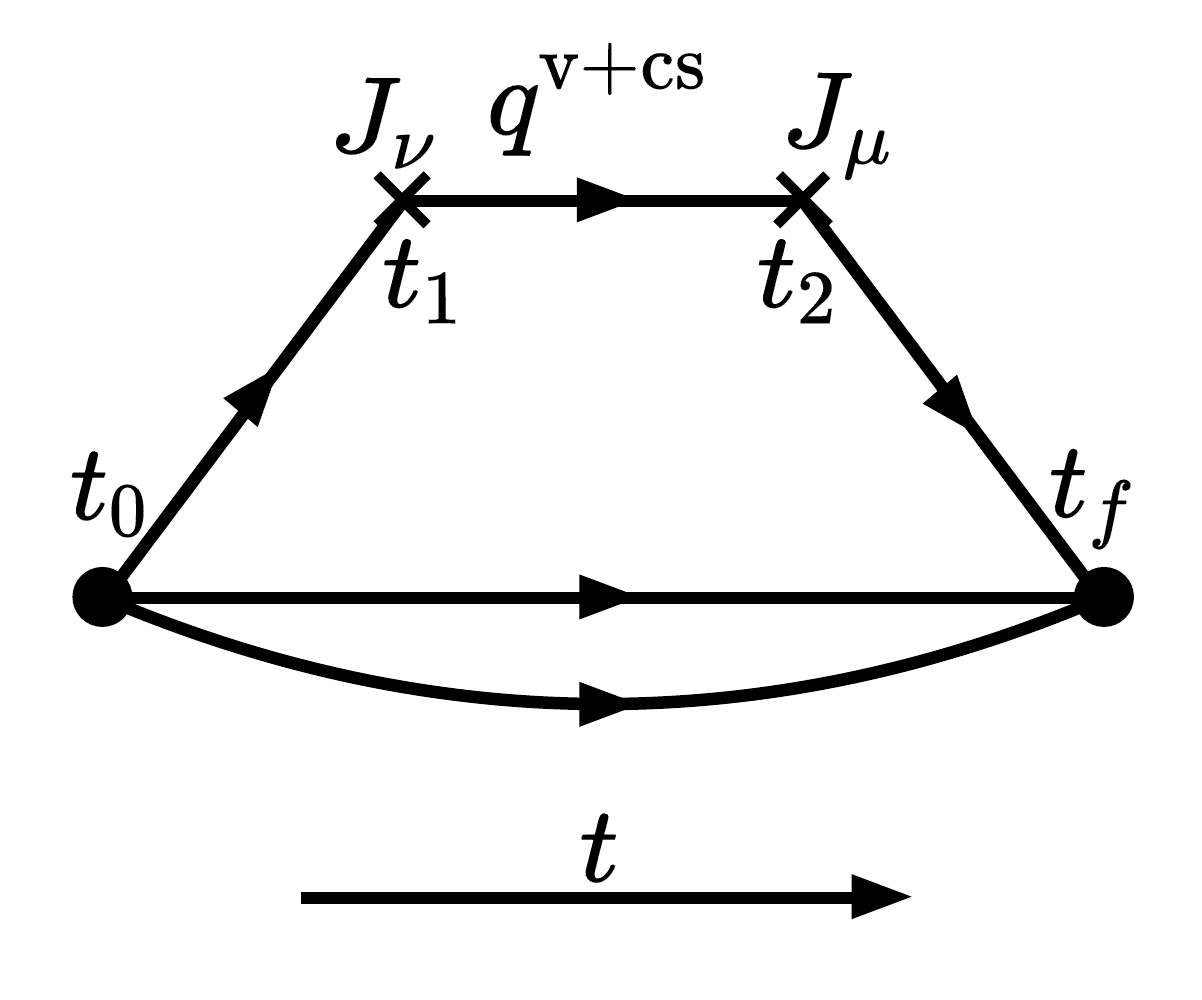}}
  \label{v+CS}}
\subfigure[]
{{\includegraphics[width=0.32\hsize]{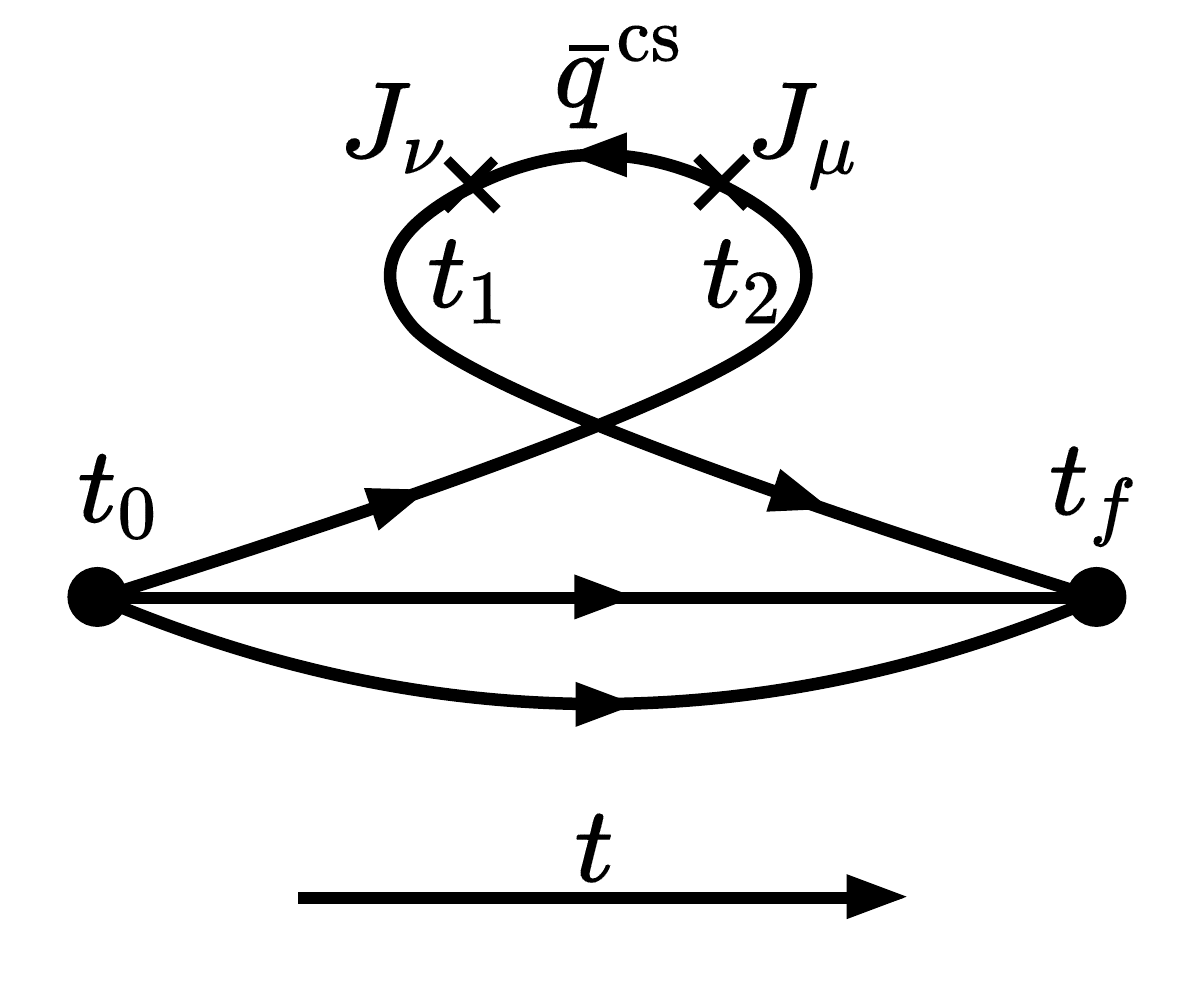}}
  \label{CS}}
%\vspace*{-0.6cm}
\subfigure[]
{{\includegraphics[width=0.32\hsize]{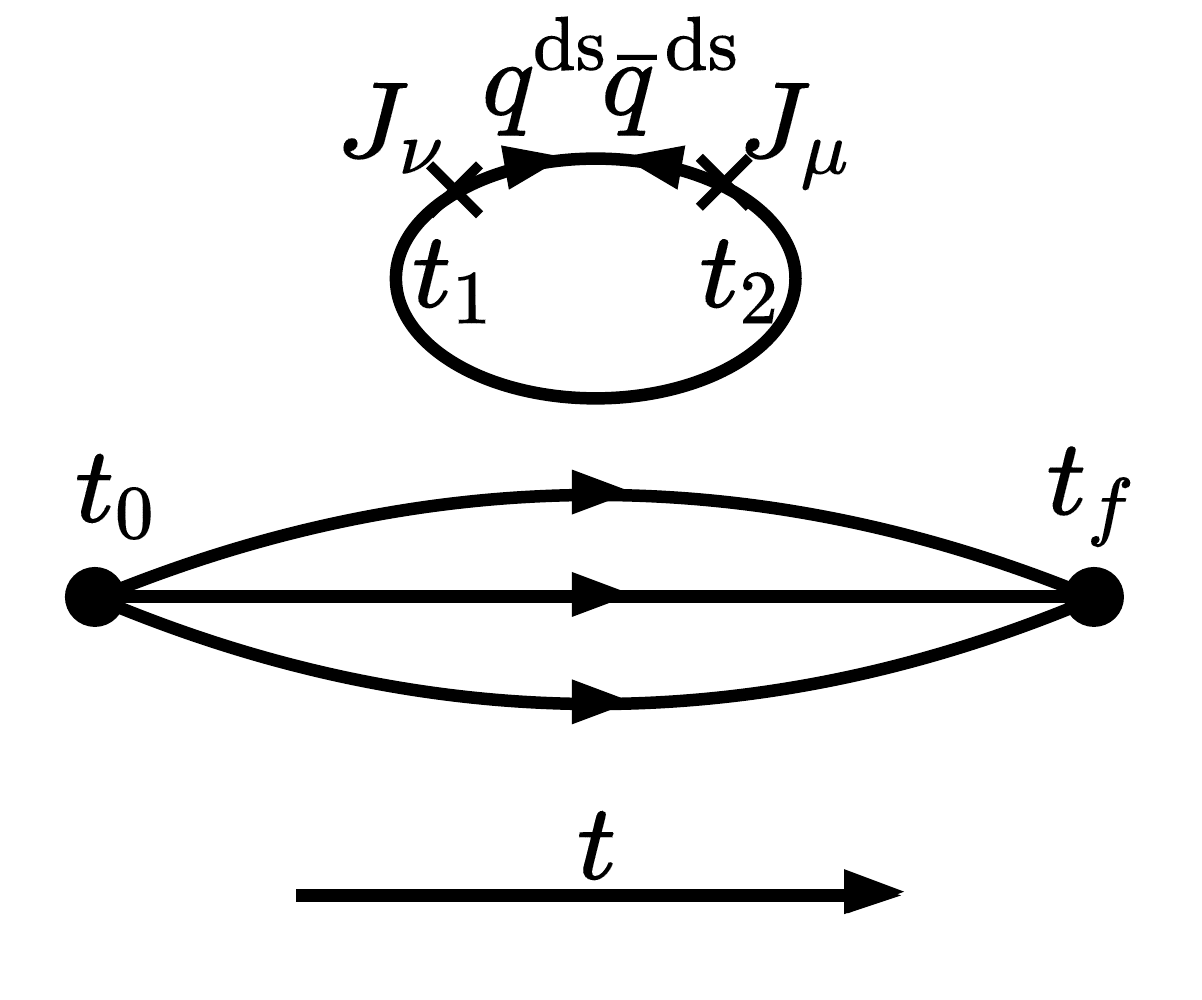}}
 \label{DS}}
 \caption{Three gauge invariant and topologically distinct insertions in the Euclidean-path integral
formulation of the nucleon hadronic tensor where the currents couple to the same quark
propagator. In the DIS region, the parton degrees of freedom are
  (a) the valence and connected sea (CS) partons $q^{v+cs}$, (b) the CS anti-partons $\bar{q}^{cs}$. Only $u$ and $d$ are present in (a) and (b) for the nucleon hadronic tensor. (c) the disconnected sea (DS) partons $q^{ds}$ and anti-partons $\bar{q}^{ds}$ with $q = u, d, s,$ and $c$.  
\label{leading-twist}}
%\newline
\vspace*{1cm}
 \subfigure[]
{{\includegraphics[width=0.32\hsize]{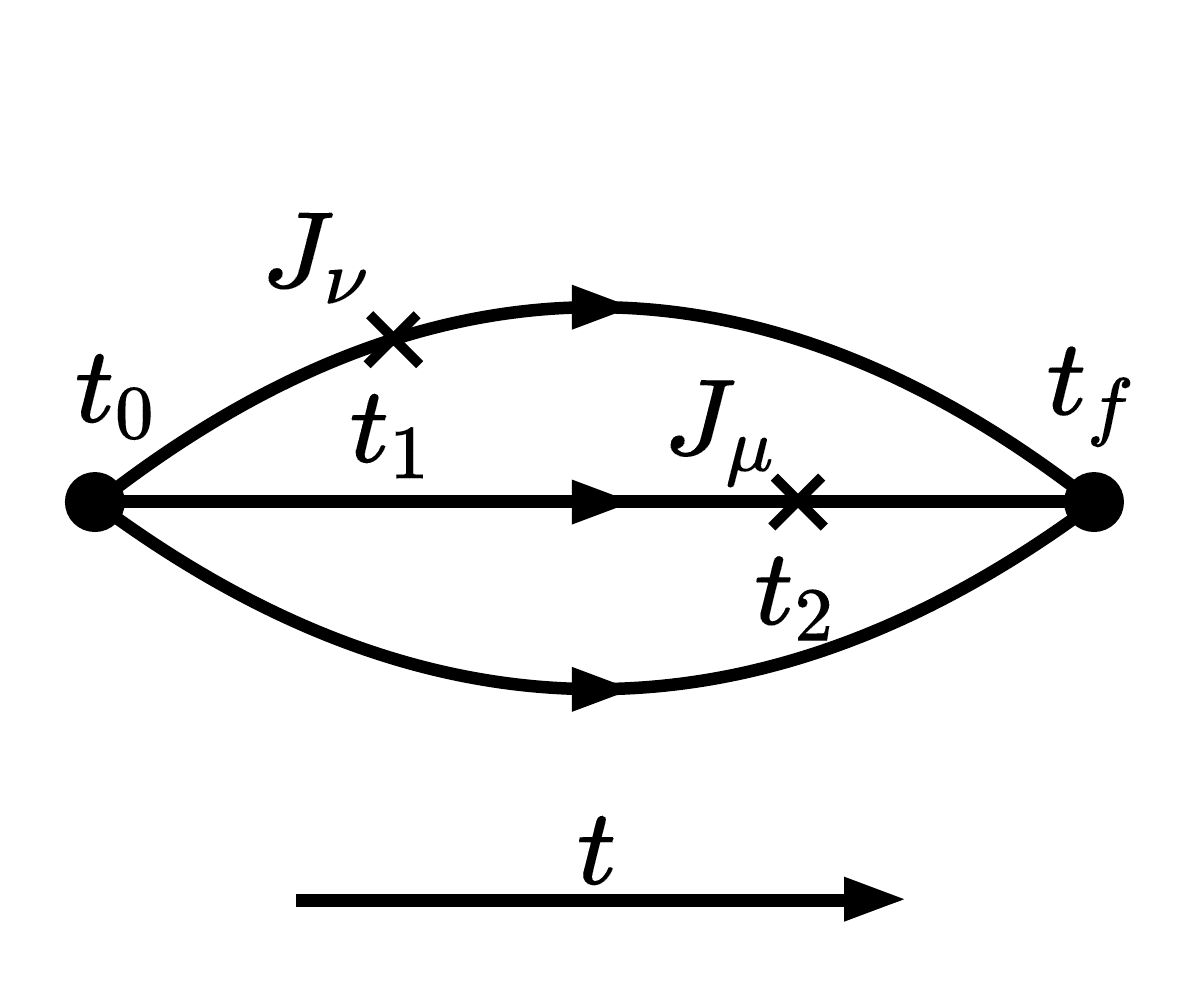}}
  \label{cat_ear_1}}
\subfigure[]
{{\includegraphics[width=0.32\hsize]{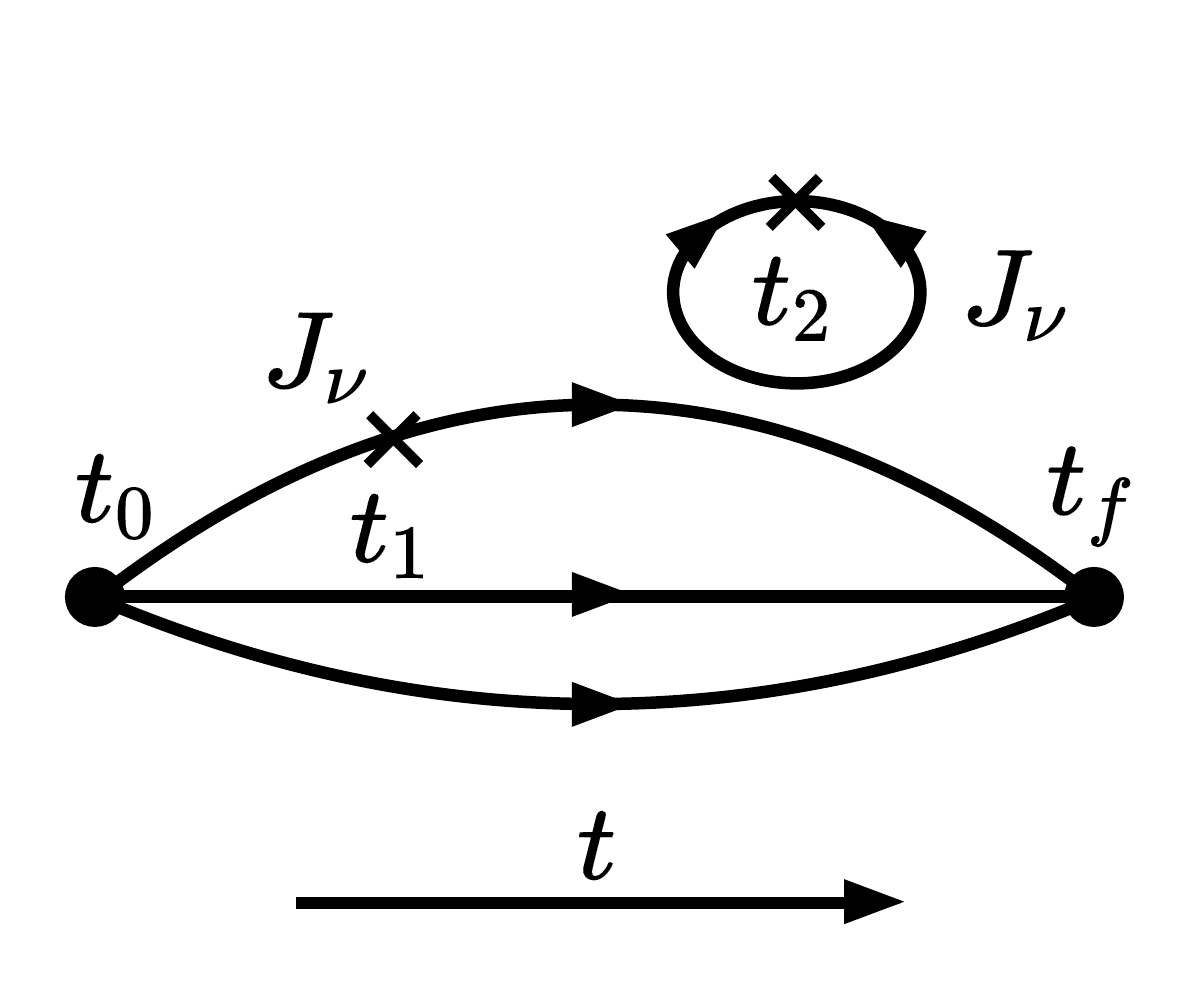}}
  \label{cat_ear_2}}
%\vspace*{-0.6cm}
\subfigure[]
{{\includegraphics[width=0.32\hsize]{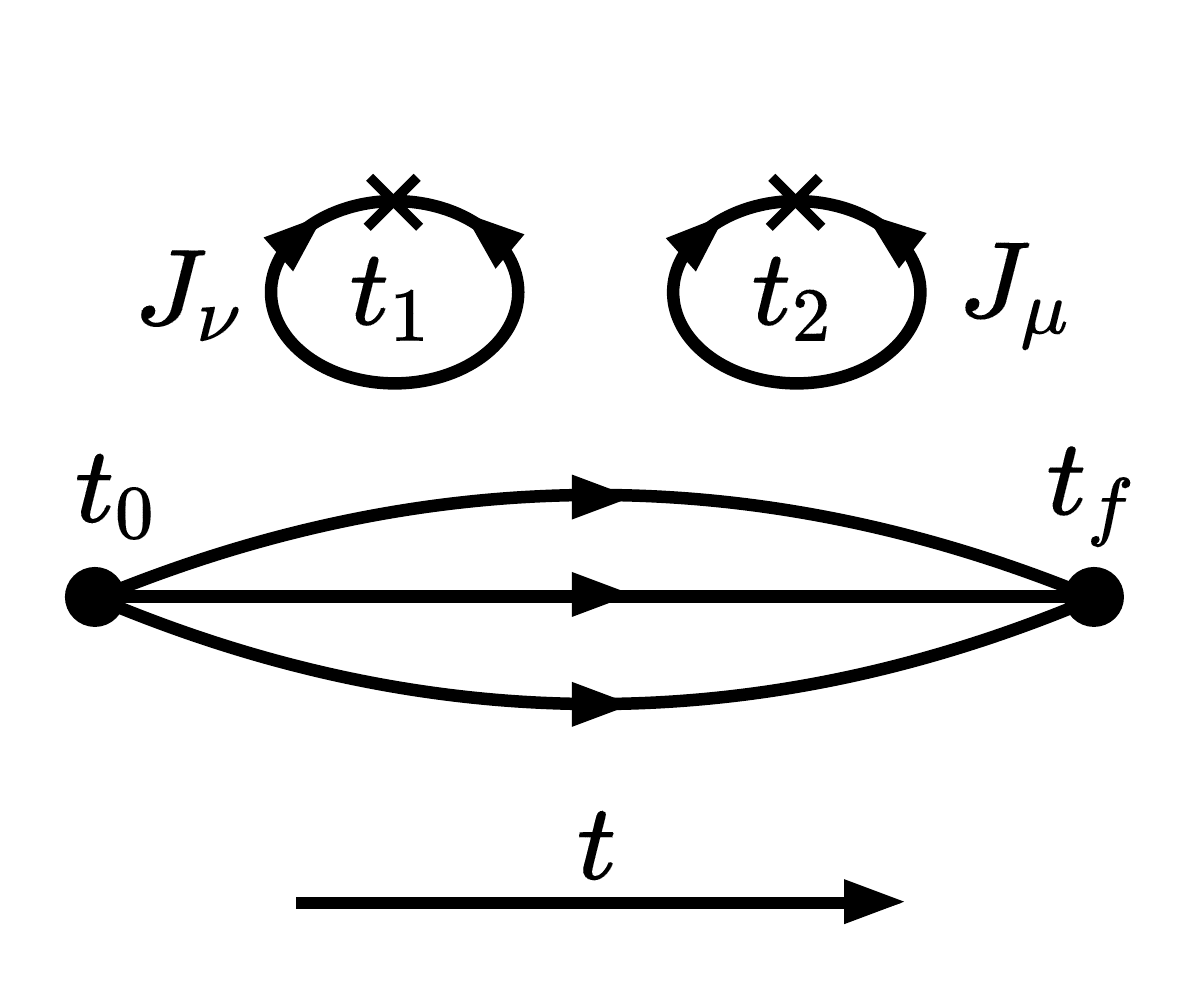}}
 \label{cat_ear_3}}
\caption{Three other gauge invariant and topologically distinct insertions where the currents are
inserted on different quark propagators. In the DIS region, they are higher-twist diagrams.
\label{higher-twist}}
\end{figure}
 It is shown~\cite{Liu:1993cv,Liu:1999ak,Liu:1998um} that, when the time ordering 
 $t_f > t_2 > t_1 > t_0$ is fixed, the 4-point function ${\rm Tr} (\Gamma_e G_{pWp}(t_0,t_1,t_2,t_f))$
 can be grouped in terms of 6 topologically distinct and gauge invariant path-integral insertions as illustrated  in Figs.~\ref{leading-twist} and \ref{higher-twist}, according to different Wick contractions among the Grassmann numbers in the two currents and the source/sink interpolation fields.  
They can be denoted as connected insertions  (CI) (Fig.~\ref{v+CS} 
Fig.~\ref{CS}, and Fig.~\ref{cat_ear_1}) where the quark lines are all connected and disconnected insertions (DI) (Fig.~\ref{DS}, Fig.~\ref{cat_ear_2}, and Fig.~\ref{cat_ear_3}) where there are vacuum polarizations associated with the currents in disconnected quark loops. Note, these diagrams depict the quarks propagating in the non-perturbative gauge background which include the fluctuating gauge fields and virtual quark loops from the fermion determinant. Only the quark lines associated with the interpolations fields and currents are drawn in these quark skeleton diagrams. 
 
 At low energy lepton-nucleon scattering, all 6 diagrams in Figs.~\ref{leading-twist} and 
 \ref{higher-twist} contribute and they are not separable.
For the elastic scattering case, the hadronic tensor $W_{\mu\nu}$ as a function of $Q^2$ is basically the product sum of the relevant nucleon form factors. For example, It is verified in a recent lattice calculation~\cite{Liang:2019frk,Liang:2020sqi} that it is the sum of Fig.~\ref{v+CS} and Fig.~\ref{cat_ear_1} that give rise to the square of the charges for the $u$ quarks in the proton in the forward limit when $J_{\mu}$ and $J_{\nu}$ are the charge current, i.e. 
$W_{44} (\vec{p} = \vec{q} = 0, \nu = 0) (u\, {\rm quark}) = (2 e_u)^2$, while the other diagrams are zero due to charge conservation. At low $\vec{q}$ and $\nu$ appropriate for a $\rho-N$ intermediate state, all connected insertions (CI) in Figs.~\ref{v+CS}, \ref{CS}, and \ref{cat_ear_1} contribute to the $\rho-N$ scattering. It is worth pointing out that Fig.~\ref{CS} includes the exchange contribution to prevent the $u/d$ quark in the loop in Fig.~\ref{DS} from occupying the same Dirac eigenstate in the nucleon propagator, enforcing the Pauli principle. In fact, Figs.~\ref{DS} and Fig.~\ref{CS} are analogous to the direct and exchange diagrams in time-ordered Bethe-Goldstone diagrams in many-body theory.

\subsection{Parton Degrees of Freedom}

In the DIS region (e.g. $Q^2 \ge 4\,{\rm GeV^2}$ and $\nu > 7$ GeV in Eq.~(\ref{nu})) as illustrated in Fig.~\ref{nu-N-spectrum}, insertions in Fig.~\ref{leading-twist} contain leading and higher twists (hand-bag diagrams), while those in Fig.~\ref{higher-twist} contain only higher twists (cat's ears diagrams). 
As far as the leading-twist DIS structure functions $F_1, F_2$ and $F_3$ are concerned,  the three diagrams in Fig.~\ref{leading-twist} are additive with contributions classified as the valence and sea partons $q^{v+cs}$ in Fig.~\ref{v+CS}, the connected sea (CS) antipartons $\bar{q}^{cs}$ in Fig.~\ref{CS}, 
and disconnected sea (DS) partons $q^{ds}$ and antipartons $\bar{q}^{ds}$ in Fig.~\ref{DS}~\cite{Liu:1993cv,Liu:1999ak,Liu:1998um}. It was pointed out that the Gottfried sum rule violation comes entirely from
the connected sea difference $\bar{u}^{cs} - \bar{d}^{cs}$ in the $F_2$ structure functions at the isospin symmetric limit~\cite{Liu:1993cv}. 

Owing to the factorization theorem~\cite{Collins:1989gx} which separates out the long-distance and short distance behaviors, the structure function $F_1$ of the hadronic tensor can be factorized as
\begin{equation} \label{factorization}
F_1 (x, Q^2) = \sum_{i } \int_x^1 \frac{dy}{y}  C_i \Big(\frac{x}{y}, \frac{Q^2}{\mu^2}, \frac{\mu_f^2}{\mu^2}, \alpha_s(\mu^2)\Big)\, f_i (y, \mu_f, \mu^2),
\end{equation}
where $i$ is summed over $q_i, \bar{q}_i, g$. $C_i$ are the Wilson coefficients and $f_i$ are the
parton distribution functions (PDFs). $\mu_f$ is the factorization scale, and $\mu$ is the renormalization scale. In practice, the global fitting programs adopt the parton degree of freedoms as $u, d, \bar{u}, \bar{d}, s, \bar{s}$ and $g$. We see that from the path-integral formalism, each of the
$u$ and $d$ have two sources, one from the connected insertion (CI) (Fig.~\ref{v+CS}) and one
from the disconnected insertion (DI) (Fig.~\ref{DS}), so are $\bar{u}/\bar{d}$ from Fig.~\ref{CS}
and Fig.~\ref{DS}. On the other hand, $s$ and $\bar{s}$ only come from the DI (Fig.~\ref{DS}).
In other words,
\begin{eqnarray}  \label{dof}
u\! &=&\! u^{v+cs} + u^{ds}, \hspace{2cm} d\,= \,d^{v+cs} + d^{ds} \nonumber \\
\bar{u}\! &=&\! \bar{u}^{cs} + \bar{u}^{ds}, \hspace{2.4cm}  \bar{d}\, = \,\bar{d}^{cs} + \bar{d}^{ds}, \nonumber \\
s\! &=&\! s^{ds},   \hspace{3.5cm} \bar{s}\, = \,\bar{s}^{ds},
\end{eqnarray}

This classification of the parton degrees of freedom (PDF) is richer than those in terms of $q$ and
 $\bar{q}$ in the global analysis in that there are two sources for the partons -- $q^{v+cs}$ and $q^{ds}$ -- and two sources for the antipartons -- $\bar{q}^{cs}$ and $\bar{q}^{ds}$. They have different small
$x$ behaviors. For connected insertion (CI) part, $q^{v+cs}, \bar{q}^{cs} \sim x^{-1/2} $ where 
$q = u,d$; whereas, for the disconnected insertion (DI) part, 
$q^{ds}, \bar{q}^{ds} \sim x^{-1} $ 
where $q = u,d,s,c$~\cite{Liu:1999ak,Liu:1998um,Liu:2012ch,Liang:2019xdx}. It is discerning to follow these degrees of freedom in moments which further 
reveal the roles of CI and DI in nucleon matrix elements. They have been intensively studied in lattice calculations which are beginning to take into account all systematic corrections.
 
 \subsection{Moments of PDFs}
 
\begin{figure}[htbp] \label{moments}
\centering
\subfigure[]
{{\includegraphics[width=0.32\hsize]{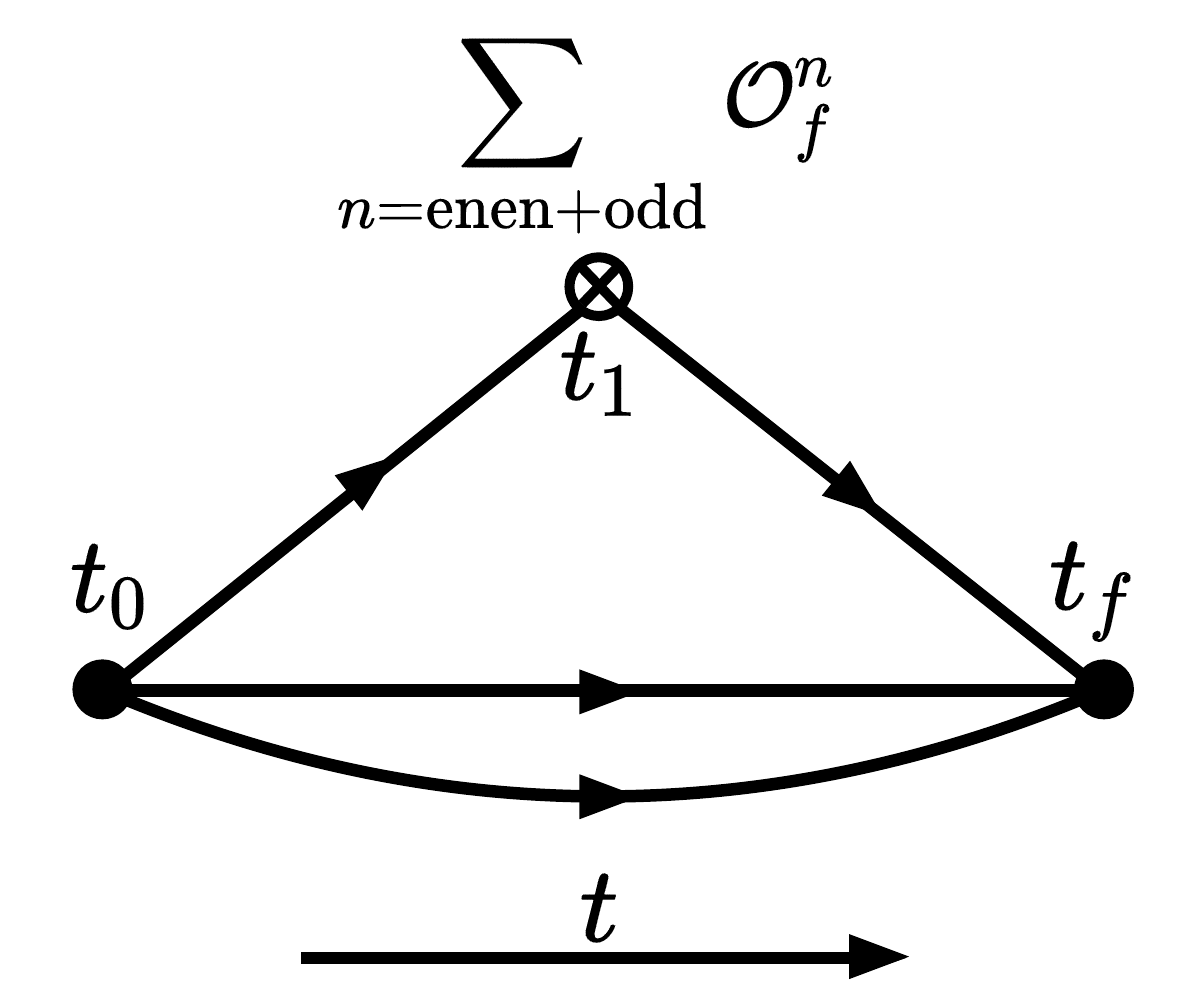}}
  \label{moment_1}}
\subfigure[]
{{\includegraphics[width=0.32\hsize]{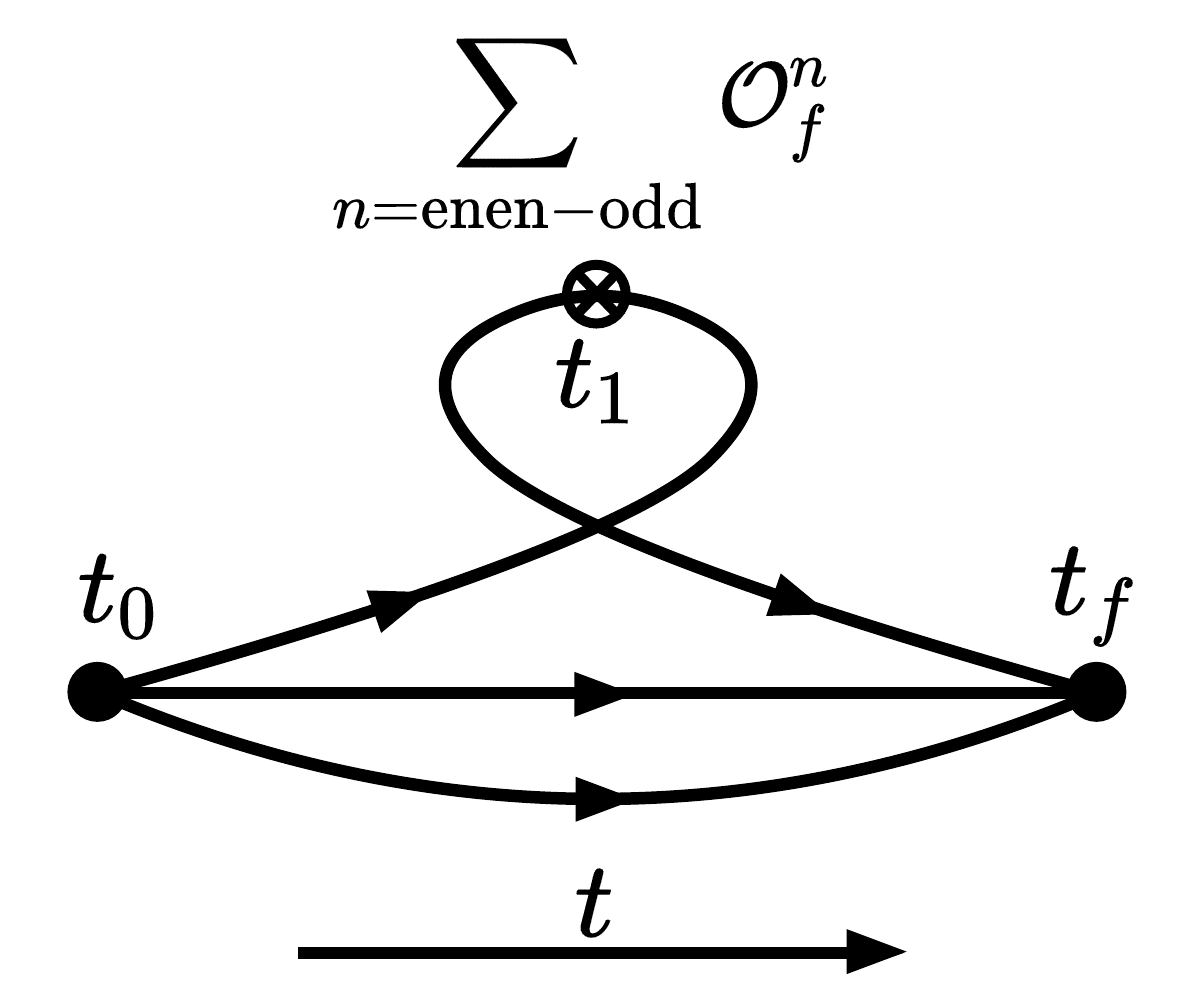}}
  \label{moment_2}}
%\vspace*{-0.6cm}
\subfigure[]
{{\includegraphics[width=0.32\hsize]{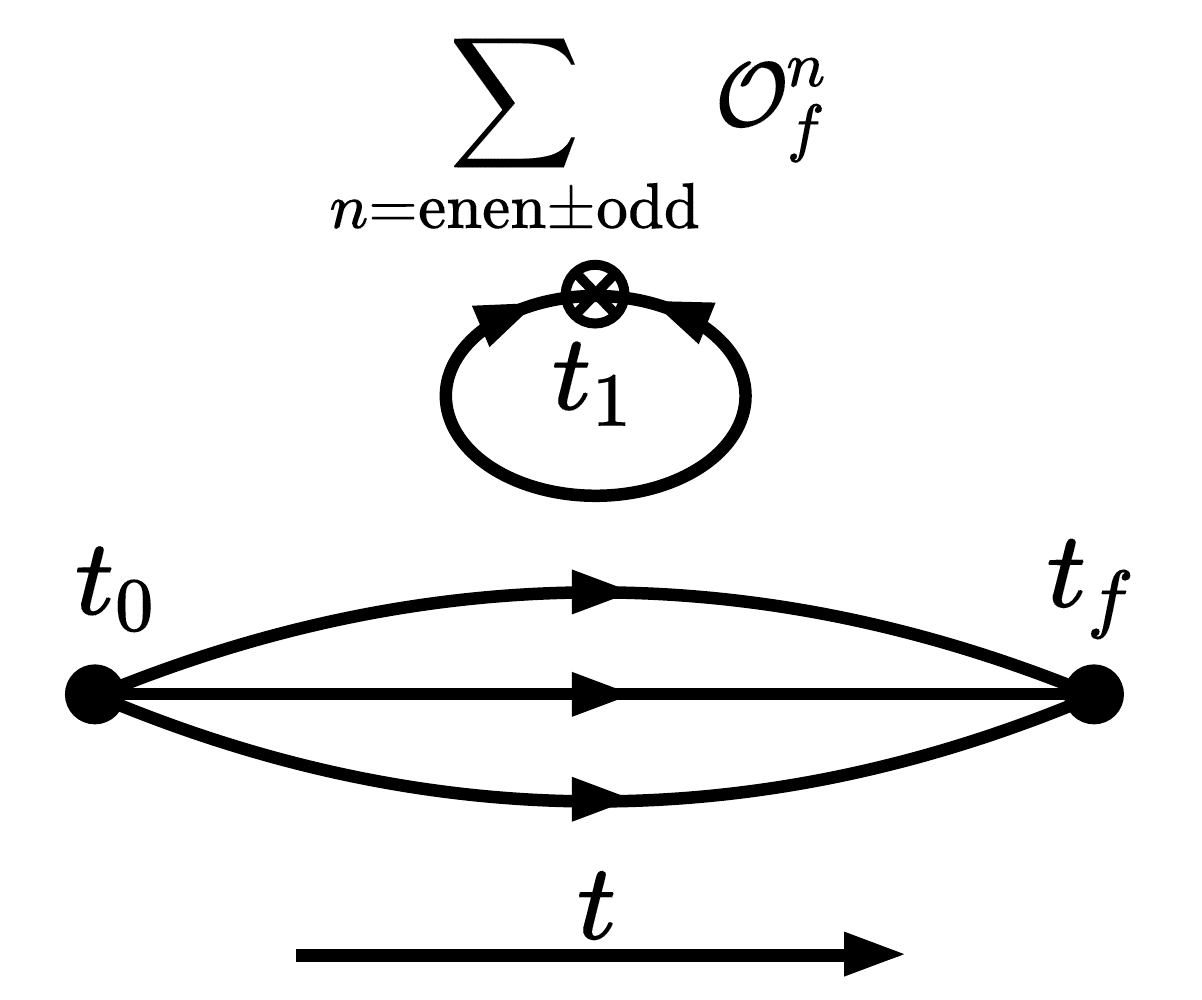}}
 \label{moment_3}}
\caption{Quark skeleton diagrams in the evaluation of matrix elements for the towers of local
operators from the short-distance expansion of $J_{\mu}(x) J_{\nu}(0)$. Figs.~\ref{moment_1},
\ref{moment_2}, and \ref{moment_3} correspond to the short distance expansions from 
Figs.~\ref{v+CS}, \ref{CS} and \ref{DS}, respectively.}
%The three-point functions after the short-distance expansion of the hadronic tensor from Fig. 1. (a) The connected insertion (CI) is derived from Fig.~\ref{val+CS} and
%ig.~\ref{CS}. (b) The disconnected insertion (DI) originates from Fig.~\ref{DS}. $\mathcal{O}_q^n$ are
%local operators which are the same as derived from OPE}
\end{figure}

 The short-distance expansion of the current-current correlator in the nucleon in 
 Eq.~(\ref{wmunu_tilde}) has been carried out~\cite{Liu:1999ak}. After applying inverse Laplace transform in Eq.~(\ref{wmunu}) 
 and dispersion relation to convert the hadronic tensor to the Minkowski space, it is shown that the total results are the same as that of the operator product expansion. However, the Euclidean path-integral formulation of the current-current correlator is composed of several components. The leading-twist forward Compton amplitude $T_{\mu\nu}$ corresponding to $q^{v+cs}$ in Fig.~\ref{v+CS}  are expanded as
 \begin{equation} \label{v+cs}
T_{\mu\nu}(q^{v+cs}) = \sum_f e_f^2 8 p_{\mu} p_{\nu}\Big\{\sum_{{\rm even},\, n =2}
\frac{(-2 q\cdot p)^{n -2}}{(Q^2)^{n -1}} A_f^n (CI) + \sum_{\rm{odd},\, n =3}
\frac{(-2 q\cdot p)^{n -2}}{(Q^2)^{n -1}} A_f^n (CI)\Big\},
\end{equation}
where $f$ indicates flavor. For the nucleon, it only involves $u$ and $d$ in
the CI. In this case,  $A_f^n (CI)$ are defined through the renormalized connected insertion (CI) matrix element at $\mu$ 
\begin{equation}   \label{matrix_element}
\langle p|\bar{\Psi} O_f^n \Psi|p\rangle_{CI} (\mu) =2\, A_f^n (CI) (\mu)\,
(p_{\mu_1}p_{\mu_2} ... p_{\mu_n} - {\rm traces}).
\end{equation}
where the renormalized operators $O_f^n$ are
\begin{equation}  \label{On}
O_f^n = Z_n(\mu)\, i \gamma_{\mu 1} (\frac{-i}{2})^{n-1} \stackrel{\leftrightarrow}
{D}_{\mu 2} \stackrel{\leftrightarrow}{D}_{\mu 3}... 
\stackrel{\leftrightarrow}{D}_{\mu n} - {\rm traces}.
\end{equation}
where $Z_n (\mu)$ is the renormalization constant. This is represented by the ratio of three-point functions illustrated in Fig.~\ref{moment_1} with the insertion of the
tower of $O_f^n$ operators and the nucleon two-point function. 

%
%The traceless $T_{\mu\nu}(q_{v+cs})$ is thus
%
%\begin{equation}  \label{qvcs}
%T_{\mu\nu}(q^{v+cs}) = \sum_{{\rm even},\, n=2} \cdots A_f^n({\rm CI}) +
%\sum_{{\rm odd},\, n=3} \cdots A_f^n({\rm CI}).
%\end{equation}   
%

 Similarly, the short-distance expansion for the CS antiparton 
in Fig.~\ref{CS} results in  a similar expression as in Eq. (\ref{v+cs}) with
the substitution $q \rightarrow - q$ and $\mu \leftrightarrow \nu$. As a result, this leads to the even 
$n$
terms minus the odd $n$ terms instead of the sum as in Eq. (\ref{v+cs}), i.e.
\begin{equation}  \label{barcs}
T_{\mu\nu}(\bar{q}^{cs}) = \sum_{{\rm even},\, n=2} \cdots A_f^n({\rm CI})  - 
\sum_{{\rm odd},\, n=3} \cdots A_f^n({\rm CI}).
\end{equation}   
In other words, the short-distance expansion of $T_{\mu\nu}$ from Fig.~\ref{CS} 
yields three-point functions with a series of insertions of the
same operators $\bar{\Psi} O_f^n \Psi$ except with a minus sign for the
odd $n$ terms. This is illustrated in Fig.~\ref{moment_2}.
By the same token, the short-distance expansion for the DS parton/antiparton 
contribution in $T_{\mu\nu}$ from Fig.~\ref{DS} gives
\begin{equation} \label{ds/bards}
T_{\mu\nu}(q^{ds}/\bar{q}_{ds}) =  \sum_{{\rm even},\, n=2} \cdots A_f^n({\rm DI}) \, \pm
\,\sum_{{\rm odd},\, n=3} \cdots A_f^n({\rm DI}).
\end{equation}
They have the same expression as $T_{\mu\nu}(q^{v+cs})$ and 
$T_{\mu\nu}(\bar{q}^{cs})$ except $A_f^n ({\rm DI})$ are from the DI part 
of the matrix element
\begin{equation}
\langle p|\bar{\Psi} O_f^n \Psi|p\rangle_{DI} = 2\, A_f^n ({\rm DI})
( p_{\mu_1}p_{\mu_2} ... p_{\mu_n} - {\rm traces}).
\end{equation}
In this case, the leading twist expansion of the DS contribution to 
$T_{\mu\nu}$ in Fig.~\ref{DS} now leads to two series of forward matrix
elements of DI. One is for $T_{\mu\nu}(q^{ds})$ with even plus odd $n$ 
terms; the other is  $T_{\mu\nu}(\bar{q}^{ds})$ with even minus odd $n$ terms as
given in Eq. (\ref{ds/bards}). Both are represented in the three-point
functions in Fig.~\ref{moment_3}. It is worth mentioning that $T_{\mu\nu}(q^{v+cs})$
in Eq. (\ref{v+cs}) and $T_{\mu\nu}(q^{ds})$ in Eq. (\ref{ds/bards}) are
the same as those derived from the contraction of the inner pair of the
quark fields in the conventional operator product expansion of the
time-ordered current-current product 
$\bar{q}(x)\gamma_{\mu}q(x)\bar{q}(0)\gamma_{\nu}q(0)$~\cite{Peskin:1995ev}.
On the other hand, $T_{\mu\nu}(\bar{q}^{cs})$  
in Eq. (\ref{barcs}) and $T_{\mu\nu}(\bar{q}^{ds})$ in Eq. (\ref{ds/bards})
are the same as those from the contraction of the outer pair of the
quark fields in the current-current product. The only difference is that
the path-integral formalism allows the separations into the CI and the DI. 

In the operator product analysis, the $A_f^n$ are the moments of PDFs
\begin{eqnarray}    
A_f^n ({\rm CI}) = M^n ({\rm CI})\equiv  \int_0^1 dx\, x^{n -1} 
(q^{v+cs}(x, \mu) + (-1)^n \bar{q}^{cs} (x, \mu)), \label{vcs}  \\         
A_f^n ({\rm DI}) = M^n ({\rm DI})\equiv \int_0^1 dx\,  x^{n -1} 
(q^{ds}(x, \mu) + (-1)^n \bar{q}^{ds}(x, \mu)).
\end{eqnarray}

When the parts in Eqs. (\ref{v+cs}), (\ref{barcs}), and (\ref{ds/bards}) are
summed up, only the even $n$ terms of the OPE are left for the vector currents $J_{\mu}$ and
$J_{\nu}$
\begin{eqnarray}  \label{sumtmu}
T_{\mu\nu}& =& T_{\mu\nu}(q_{v+cs}) + T_{\mu\nu}(\bar{q}_{cs}) 
+ T_{\mu\nu}(q_{ds} + \bar{q}_{ds})   \nonumber \\
 &=& 2 \sum_{{\rm even}, \, n=2} \cdots (A_f^n ({\rm CI}) + A_f^n ({\rm DI})).
\end{eqnarray}
This is the same as that derived from the ordinary OPE. However, 
what is achieved with the path-integral formulation is the separation 
of CI from DI, in addition to the separation of 
partons from antipartons. This separation facilitates  the identification of
the CS parton as the source of the Gottfried sum rule violation~\cite{Liu:1993cv}, and an
extended set of evolution equations for separate $\bar{q}_{cs}(x, Q^2)$ and 
$\bar{q}_{ds}(x, Q^2)$~\cite{Liu:2017lpe}.
We should emphasize that, for a given moment, there is no more distinction between
$q^{v+cs}$ and $\bar{q}^{cs}$, nor with $q^{ds}$ and $\bar{q}^{ds}$. There are
only CI and DI matrix elements.

\subsection{Renormalization and evolution}  \label{ren-evo}

The SU(3) flavor dependence of the moments is usually expressed in terms of
iso-vector, flavor octet and flavor singlet. Here in the path-integral formulation, there
are more moments that can be -- and have been -- evaluated on the lattice. For example,
the renormalization matrix for the renormalized second moment $\langle x \rangle$ has the following structure~\cite{Yang:2018nqn,Liang:2018pis}
\begin{equation}
\left(\begin{array}{c}
%\Delta u^{\overline{\rm MS}}({\rm CI})\\
%\Delta d^{\overline{\rm MS}}({\rm CI})\\
\langle x\rangle_ u ({\rm CI})^R(\mu)\\
\langle x\rangle_ d ({\rm CI})^R(\mu)\\
\langle x\rangle_ u ({\rm DI})^R(\mu)\\
\langle x\rangle_ d ({\rm DI})^R(\mu) \\
\langle x\rangle_ s({\rm DI})^R(\mu) \\
\langle x\rangle_ G({\rm DI})^R (\mu)
\end{array}\right)=\left(\begin{array}{cccccc}
Z_C & 0 & 0 & 0 &0 &0 \\
0 & Z_C & 0 & 0 & 0 & 0\\
Z_D & Z_D & Z_C +Z_D & Z_D& Z_D &Z_{qG} \\
Z_D & Z_D & Z_D & Z_C+Z_D& Z_D &Z_{qG} \\
Z_D & Z_D & Z_D & Z_D & Z_C + Z_D &Z_{qG} \\
Z_{Gq} & Z_{Gq} & Z_{Gq} & Z_{Gq} & Z_{Gq}  &Z_{GG} \\
\end{array}\right)
\left(\begin{array}{c}
\langle x\rangle_ u ({\rm CI}) \\
\langle x\rangle_ d ({\rm CI})\\
\langle x\rangle_ u ({\rm DI})\\
\langle x\rangle_ d ({\rm DI})\\
\langle x\rangle_ s({\rm DI}) \\
\langle x\rangle_ G({\rm DI})
\end{array}\right) \label{eq:renorm_6}
\end{equation}
After linear combinations of CI and DI matrix elements in the equations in 
Eq.~(\ref{eq:renorm_6}), one can reduce
them to those in the flavor-SU(3) representation
\begin{equation}
\left(\begin{array}{c}
%\Delta u^{\overline{\rm MS}}({\rm CI})\\
%\Delta d^{\overline{\rm MS}}({\rm CI})\\
\langle x\rangle_ {u-d}^R(\mu)\\
\langle x\rangle_ {u+d - 2s}^R(\mu)\\
\langle x\rangle_ {u+d+s}^R(\mu)\\
\langle x\rangle_ G^R (\mu)
\end{array}\right)=\left(\begin{array}{cccccc}
Z_C & 0 & 0 & 0  \\
0 & Z_C & 0 & 0 \\
0 & 0 & Z_C + N_f Z_D & N_f Z_{qG} \\
0 & 0 & Z_{Gq}   &Z_{GG} \\
\end{array}\right)
\left(\begin{array}{c}
\langle x\rangle_ {u-d} \\
\langle x\rangle_ {u+d -2s}\\
\langle x\rangle_ {u+d+s} \\
\langle x\rangle_ G
\end{array}\right) \label{eq:renorm_4}
\end{equation}

It should be noted that there are 6 observables in Eq.~(\ref{eq:renorm_6}) 
(5 if one assumes isospin symmetry so that 
$\langle x\rangle_ u ({\rm DI}) = \langle x\rangle_ d ({\rm DI})$), while there are
4 in Eq.~(\ref{eq:renorm_4}). This means that, by separating the CI and DI, the path-integral 
formulation has more information than that of the flavor classification. Since the lattice
calculations are organized with separate CI and DI moments, it is natural to ask how to
accommodate such a separation in the global analysis so that one can make a one-to-one 
comparison between the lattice results and those from the global fits.
At the moment, the CS partons and DS partons are not separated in global analyses, one could
not make such a comprehensive comparison and is only limited to the isovector and strangeness
quantities. 

Attempts have been made~\cite{Liu:2012ch,Peng:2014uea,Peng:2014eza,Liang:2019xdx} to separate out the CS and DS partons by combining strange parton
distribution from a HERMES experiment~\cite{Airapetian:2008qf}, $\bar{u} + \bar{d}$ from
the CT10 analysis~\cite{Lai:2010vv}, and the ratio $\langle x\rangle_{s} / \langle x\rangle_ u ({\rm DI})$ from lattice calculations~\cite{Doi:2008hp,Liang:2019xdx}. 

It has been pointed out~\cite{Liu:2017lpe} that in order to have a direct one-to-one comparison with the
11 parton distributions in Eq.~(\ref{dof}) (including the glue distribution) from the hadronic tensor
which can be calculated on the lattice~\cite{Liu:2016djw,Liang:2017mye,Liang:2019frk}, 
the NNLO evolution equations need to be extended to accommodate these partons. In certain
cases, such as the moments~\cite{Liang:2019frk}, precise lattice calculations can be used to help constrain the global PDF analysis and the small $x$ behavior. The extended evolution equations 
are worked out~\cite{Liu:2017lpe} for the 11 parton distributions in Eq.~(\ref{dof}) with an explicit
separation of CS and DS. Due to the linear nature of the evolution equations, these equations
can be combined and reduced to the present DGLAP equations with $\bar{q}^{cs}$ and $\bar{q}^{ds}$
merged into $\bar{q}$. But, only through the fully separated CS and DS degrees of freedom in the
extended evolution equations can the CS and DS be separated at different $Q^2$~\cite{Liu:2017lpe}. 
This aspect is essential for the global analysis of PDF to fit experimental data at different $Q^2$.
 
Another aspect of the separation of the CI from DI is shown in Eq.~(\ref{eq:renorm_6}) where the
zeros reflect the fact that the CI can mix into glue and DI, but not vice versa. There are
no valence like partons in the strange and glue partons. This is reflected in the extended evolution
equations~\cite{Liu:2017lpe} and it turns out to be crucial in resolving the puzzles presented in 
introduction in Sec.~\ref{intro} as we shall see later.

\section{Feynman-$x$ Approach}    \label{Feynman-x}

\begin{figure}[htbp]
\centering
\subfigure[]
{{\includegraphics[width=0.4\hsize]{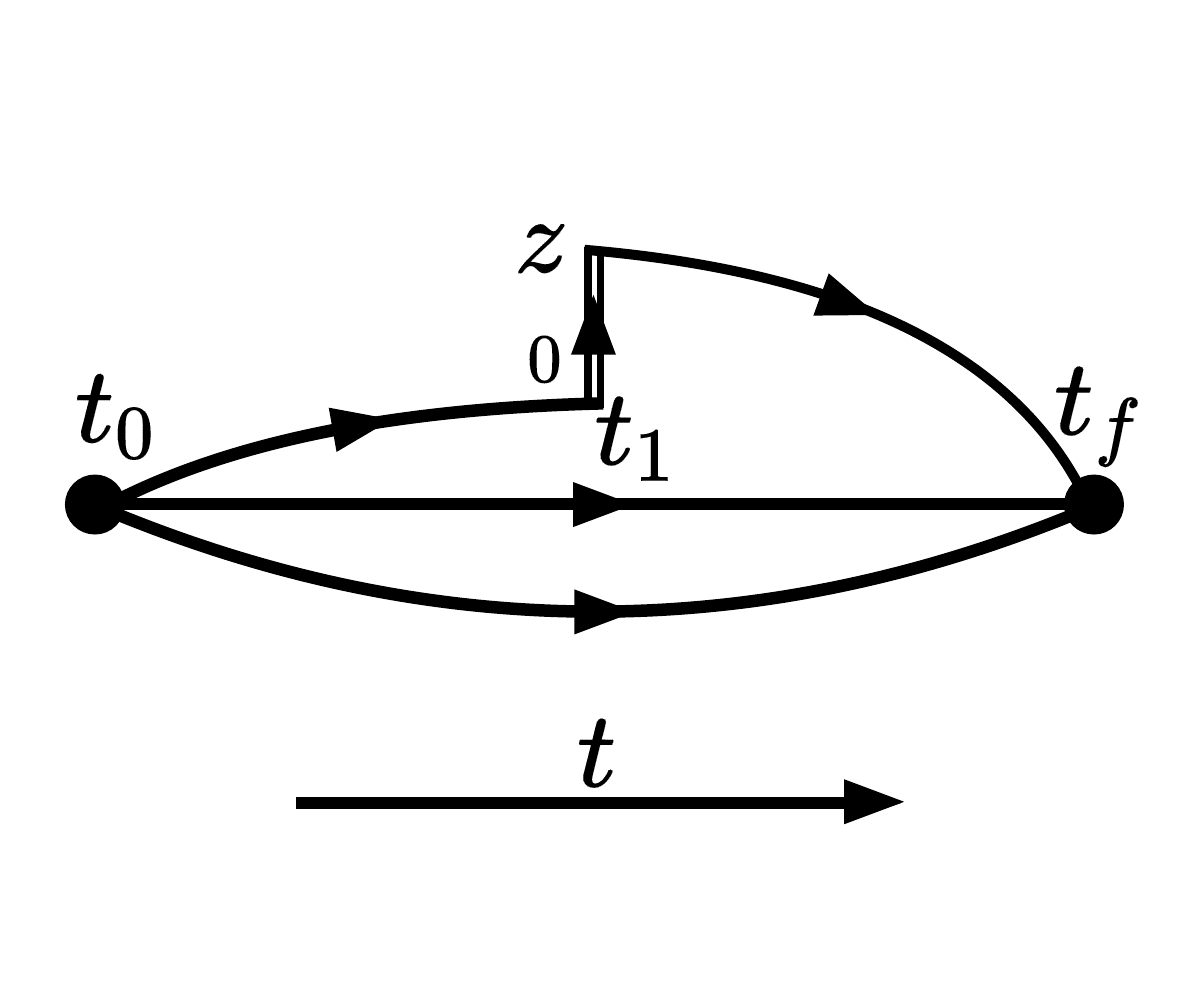}}
  \label{QPDF_CI}}
\subfigure[]
{{\includegraphics[width=0.4\hsize]{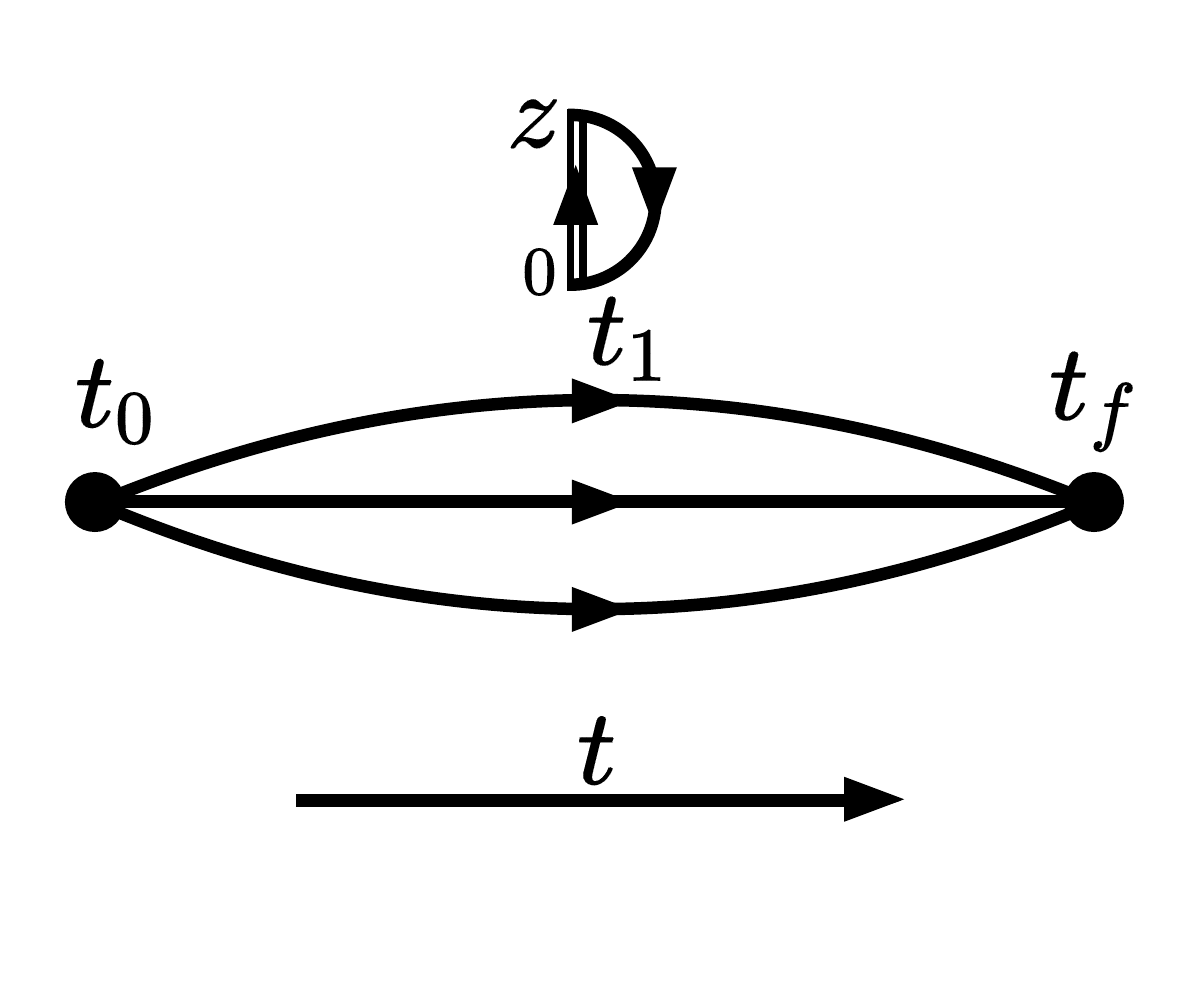}}
  \label{QPDF_DI}}
%\vspace*{-0.6cm}
\caption{Path-integral diagrams of quasi-PDF as defined in Eq.~(\ref{qpdf}). 
The double lines represent the Wilson line operators. There are two diagrams -- (a) for CI
and (b) for DI, respectively.}
 \label{QPDF}
 \end{figure}

        A number of approaches have been developed in recent years to calculate the PDFs in 
Feynman-$x$ with lattice calculation~\cite{Ji:2013dva,Radyushkin:2017cyf,Ma:2017pxb}. It was first pointed out by Ji~\cite{Ji:2013dva} that one can approximate a light-cone PDF by boosting a time independent quasi PDF defined on the lattice to a large momentum frame with an expansion in powers of the inverse hadron momentum. At leading twist, the boosted quasi PDF can be renormalized in the $\overline{\rm{MS}}$ scheme and factorized into the light-cone PDF and a perturbative matching coefficient. The matching coefficient can be calculated via the
large momentum effective theory (LaMET)~\cite{Ji:2014gla}. The unpolarized quasi-PDF is defined on the lattice
\begin{equation}  \label{qpdf}
\tilde{q} (x, P_z) \equiv \int_{-\infty}^{\infty} \frac{dz}{4\pi} e^{i x\,P_z\,z} \langle P|\bar{\psi}(z)\, \Gamma \,U(z,0)\, \psi(0)|P\rangle,
\end{equation}
where $\Gamma = \gamma_4$ or $\gamma_z$ and the Wilson line is
\begin{equation}
U(z,0) = P\, exp\,\big(-ig \int_0^z dz' A_z(z')\big).
\end{equation}
There are two path-integral diagrams associated with the the four-point function in Eq.~(\ref{qpdf}).
As we can see in Fig.~\ref{QPDF}, they are again separated into  connected insertion (CI) and
disconnected insertion (DI) as in the Euclidean formulation of the hadronic tensor in Sec.~\ref{HTparton}. 
Unlike the PDF defined on the light-cone which is boost invariant, the quasi-PDF in Eq.~(\ref{qpdf}) depends on the nucleon momentum $P_z$.  When $P_z$ is much larger than the nucleon mass
M and $\Lambda_{\rm{QCD}}$, the quasi-PDF in the CI can be factorized into a matching coefficient and the PDF as proved in the diagrammatic approach~\cite{Ma:2014jla} and via the operator product 
expansion~\cite{Izubuchi:2018srq}
\begin{equation}.  \label{factor-quasi}
\tilde{q} (x, \frac{\mu}{P_z}) = \int_{-1}^{1} \frac{dy}{|y|}\, C\,\Big( \frac{x}{y}, \frac{\mu}{|y|P_z}
\Big)\, q(y, \mu) + \mathcal{O}\Big(\frac{M^2}{P_z^2}, \frac{\Lambda_{\rm{QCD}}^2}{x^2\,P_z^2}\Big),
\end{equation}
where $\mu$ is the factorization scale, and the power corrections are in terms of 
$M^2/P_Z^2$~\cite{Chen:2016utp} and $\Lambda_{\rm{QCD}^2}/x^2\,P_z^2$~\cite{Ma:2014jla}. It also shows that the factorization formula for the quasi-PDF has $p_z = |y| P_Z$ in the coefficient function $C$. As such, there are no moment relations between the light-cone PDF and the quasi-PDF.

%We note that the target PDF $q(y, \mu)$ is inside the integral in Eq.~(\ref{factor-quasi}), similar to
%the cases of global analysis in Eq.~(\ref{factorization}) and the hadronic tensor in Eq.~(\ref{Laplace}).
%They all involve an inverse problem. Like the global analysis, one can assume a functional form for
%$q(y, \mu)$ and fit the lattice data of the quasi-PDF $\tilde{q} (x, \frac{\mu}{P_z})$.

We show two recent lattice calculations of quasi-PDF in the CI from Fig.~\ref{QPDF_CI}.
\begin{figure}[htbp]     
\centering
\subfigure[]
{{\includegraphics[width=0.445\hsize]{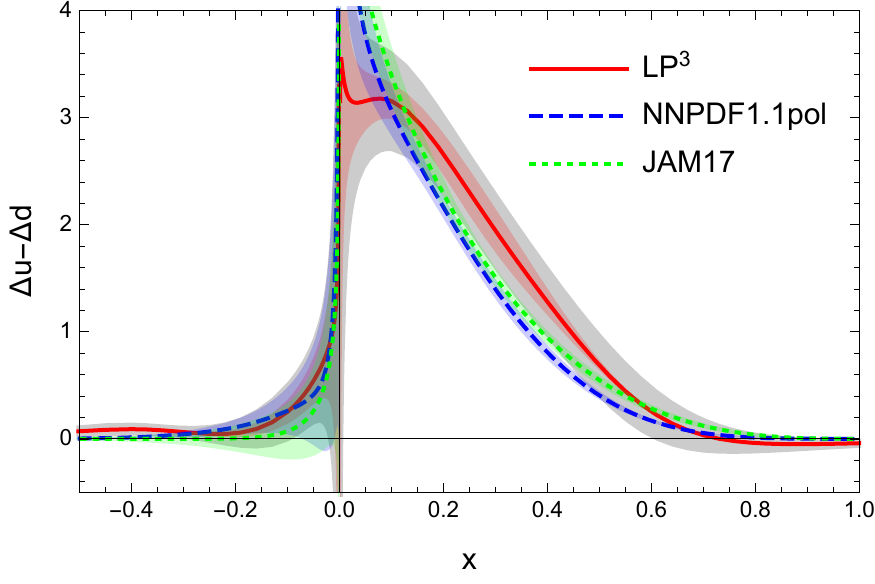}}
  \label{QPDF_LP3}}
\subfigure[]
{{\includegraphics[width=0.4\hsize]{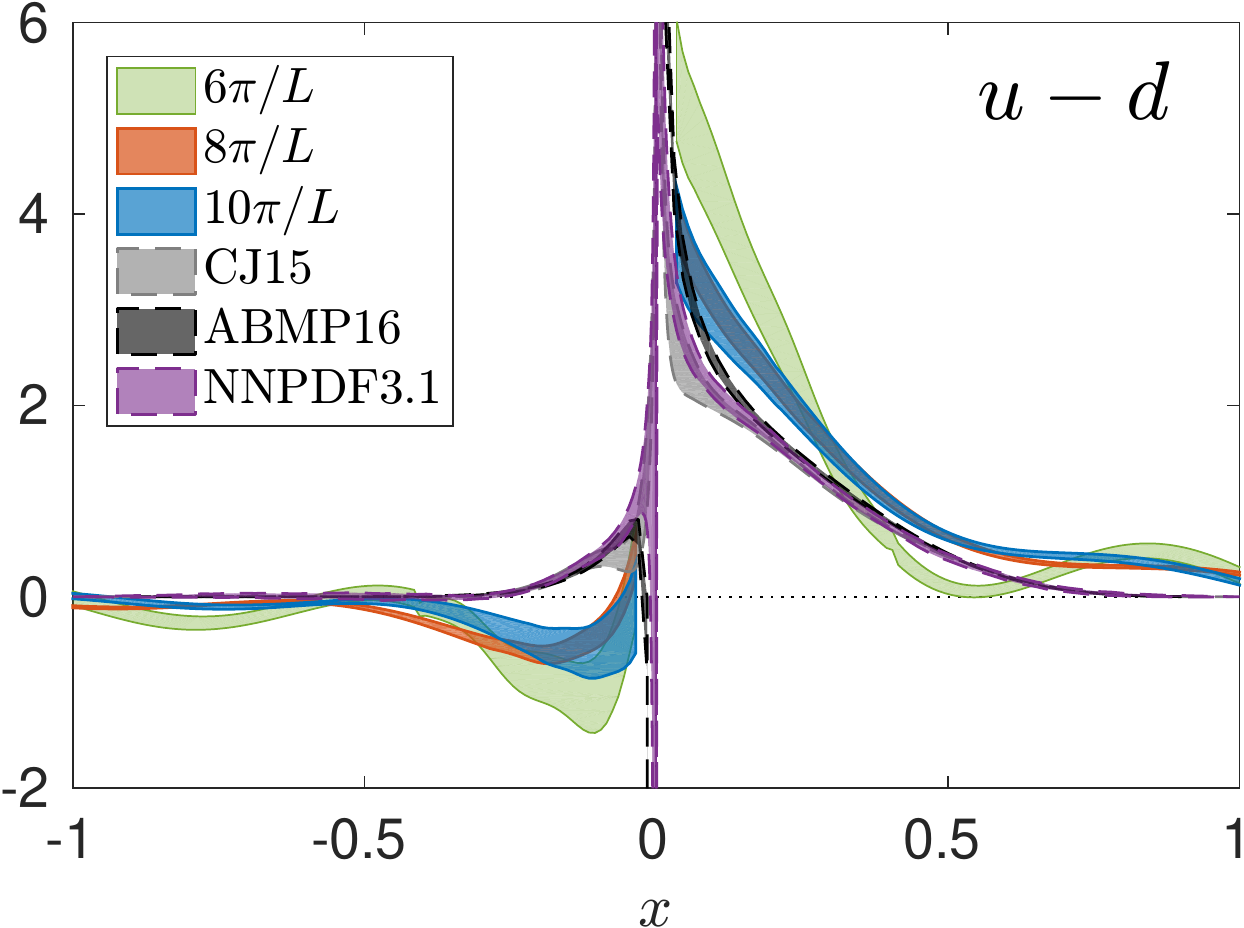}}
  \label{QPDF_ETMC}}
%\vspace*{-0.6cm}
\caption{Lattice calculation of quasi-PDF with LaMET for the connected insertion (CI).
 (a) is the polarized $\Delta u - \Delta d$ from LP3 at $\mu = 3$ GeV and (b) is 
 the unpolarized $u -d$ from ETMC at $\mu = 2$ GeV.
  \label{LatticeQPDF}}
  \end{figure}
Fig.~\ref{QPDF_LP3}  shows the calculation of the polarized PDF ($\Delta u - \Delta d$)
on a lattice with lattice spacing $a = 0.09$ and $\mu = 3$ GeV fm from the LP3 collaboration~\cite{Lin:2018pvv}  and Fig.~\ref{QPDF_ETMC}  shows the unpolarized PDF ($u -d$) from a lattice with $a = 0.0938$ fm and 
$\mu = 2$ GeV from the ETMC collaboration~\cite{Alexandrou:2018pbm}. Both of them are calculated at the physical pion mass and with one-loop matching from LaMET. To better control the errors at
small x with higher momenta and excited states, variational approaches~\cite{Chen:2005mg} can be employed.

Since the light-cone parton distribution has support $-1 \le x \le 1$ with large enough $P_Z$, it
is different from those defined from the hadronic tensor which has support $0 \le x \le 1$.
The light-cone PDFs with $x> 0$ correspond to partons from the hadronic tensor through the
factorization in Eq.~(\ref{factorization}) and those with $x <0$ correspond to antipartons
from the hadronic tensor with a negative sign. As we mentioned in Sec.~\ref{ren-evo},  partons
in the CI have no mixing from the DS and gluons. Therefore, at the same $\overline{\rm MS}$ scale 
$\mu$, the CI light-cone PDFs are identified as those from the hadronic tensor, i.e.
\begin{eqnarray}   \label{qHT}
q(x > 0, \mu) ({\rm CI})&=& q^{v+cs}(x, \mu), \nonumber \\
q(x < 0, \mu) ({\rm CI}) &=& -\, \bar{q}^{cs} (|x|, \mu).
\end{eqnarray}
It is clear from Fig.~\ref{LatticeQPDF} that the CS antipartons in the region $x < 0$ are explicitly revealed.  Similarly, one can make the corresponding identifications for the DS partons and antipartons in Eq.~(\ref{dof}).
 
 Another way to show that parton degrees of freedom in quasi-PDF with LaMET and
those  from the hadronic tensor are the same is to look at the operator product expansion. 
After renormalization to remove the power divergence from the Wilson line self-energy,
the renormalized $\bar{\psi(z)} \Gamma U(z,0) \psi(0)$ operator can be expanded in terms
of the gauge-invariant operators as $z^2 \longrightarrow 0$
\begin{equation}
\bar{\psi(z)} \Gamma U(z,0) \psi(0)_R =  \sum_{n=0}^{\infty}  C_n(\mu^2 z^2) \frac{(-iz)^n}{n!}
e_{\mu_1} ... e_{\mu_n} O^n (\mu),
\end{equation}
where $O^n (\mu)$ is that given in Eq.~(\ref{On}) and its matrix element defined in
Eq.~(\ref{matrix_element}), with $A_f^n$ being the $n^th$ moments of the DPFs. In terms of
the distribution on the light-cone, $A^n$ for the $u$ and $d$ flavors in the CI are~\cite{Izubuchi:2018srq} 
\begin{equation}  \label{moment-lc}
A^n (\mu) ({\rm CI}) = \int_{-1}^{1} dx\, x^{n-1} q (x, \mu) ({\rm CI})
\end{equation}
In terms of the PDFs defined from the hadronic tensor in Eq.~(\ref{qHT})
\begin{equation}
A^n (\mu) ({\rm CI}) = \int_0^1 dx\, x^{n-1} \Big( q^{v+cs}(x) - (-1)^{n-1} \bar{q}^{cs}(x)\Big).
\end{equation}
This is the same as Eq.~(\ref{vcs}). One can further use inverse Mellin transform to find
$q^{v+cs}$ and $\bar{q}^{cs}$ separately. One can carry out the inverse Mellin transform 
for $A^n$ for even and odd n and obtain
\begin{eqnarray}
q^{v+cs}(x) = \int_{c-i\infty}^{c+i\infty} dn\, x^{-n} A^n (n= {\rm even})  + \int_{c-i\infty}^{c+i\infty} dn\, x^{-n} A^n (n={\rm odd}) \\
\bar{q}^{cs}(x) = \int_{c-i\infty}^{c+i\infty} dn\, x^{-n} A^n (n= {\rm even}) -  \int_{c-i\infty}^{c+i\infty} dn\, x^{-n} A^n (n= {\rm odd})
\end{eqnarray}
They have the same even-odd patterns as those in Eqs.~({\ref{v+cs}) and (\ref{barcs}).
Similarly, one can show that in the DI, the $x> 0/x <0$ distribution is $q^{ds}/\bar{q}^{ds}$.
There are other Feynman-$x$ approaches to calculating PDFs on the lattice, such as
pseudo-PDF by Radyushkin~\cite{Radyushkin:2017cyf} and lattice cross-section by
Ma and Qiu~\cite{Ma:2017pxb}. After renormalization and matching to the light-cone, any
valid approach will have the CI moments expressed in Eq.~(\ref{moment-lc}) and their
parton degrees of freedom are also identified with those from the hadronic tensor.

As of now, all the lattice calculaitons with the Feynman-x approaches report the $u-d$ combination
in the CI. As we have shown in Sec.~\ref{ren-evo} that the glue and DI do not mix into the CI. 
Therefore, it is just as meaningful to report the CI $u$ and $d$ distributions separately. 

\section{Resolving the puzzles of the valence parton definition} \label{resolution}

     Once we have identified the parton degrees of freedom from the path-integral formalism of QCD,
 we are ready to answer the puzzles about the definition of the valence in Sec.~\ref{intro}. 
 First of all, the quark loops in Figs.~\ref{DS} or \ref{QPDF_DI} have a separate flavor trace from that 
 involving the interpolation field ($uud$ for the proton in this case) for the nuclear propagator, the partons in the loops do not have any knowledge of the content of the interpolation field and are, thus,
not part of the valence. Therefore, the valence should be defined from the CI in Figs.~\ref{v+CS} and \ref{CS}, i.e.
\begin{equation}   \label{lattice-valence}
q^v \equiv q^{v+cs} - \bar{q}^{cs}
\end{equation} 
Since the strange and charm partons do not have CI contributions, they are not part of the
valence. The question whether the strange is part of the valence contribution as posted in Sec.~\ref{intro} is answered by Eq.~(\ref{lattice-valence}). 

Regarding question 2 in Sec.~\ref{intro}, we see that the phenomenologically  defined $q^{-}$ is 
not the valence $q^v$. Rather it also contains DS partons 
 \begin{equation} \label{q-lattice}
 q^- \equiv q - \bar{q} = q^{v+cs} - \bar{q}^{cs} + q^{ds} - \bar{q}^{ds} = q^v + q^{ds} - \bar{q}^{ds}.
 \end{equation}  

 The NNLO evolution equations with separate CS and DS have been developed for 
 a maximum of 11 parton degrees of freedom, namely $u^{v+cs}, \bar{u}^{cs}, u^{ds}, \bar{u}^{ds},
 d^{v+cs}, \bar{d}^{cs}, d^{ds}, \bar{d}^{ds}, s, \bar{s}$ and $g$~\cite{Liu:2017lpe}. 
In view of the fact that $q^-$ is made of $q^v$ and $q^{ds} - \bar{q}^{ds}$, we find that
that Eq.~(\ref{q-}) is in fact a linear combination of two equations,
 \begin{eqnarray}
 \frac{dq_i^v}{dt} \!\!&=&\!\!  (P_{qq}^c - P_{q\bar{q}}^c) \otimes  q_i^v     \label{val}  \\
 \frac{d (q_i^{ds} - \bar{q}_i^{ds})}{dt}   \!\! &=& \!\!   (P_{qq}^d - P_{q\bar{q}}^d) \otimes  \sum_k (q_k^{ds}- \bar{q}_k^{ds})   +  (P_{qq}^d - P_{q\bar{q}}^d) \otimes \sum_k q_k^v     \label{ds-diff} \\
 \end{eqnarray} 
where we have used $P_{ii}^c = P_{\bar{i}\bar{i}}^c = P_{qq}^c$, 
$P_{i\bar{i}}^c = P_{\bar{i}i}^c = P_{q\bar{q}}^c$, $P_{ik}^d = P_{\bar{k}\bar{i}}^d = P_{qq}^d$, and $P_{i\bar{k}}^d = P_{\bar{i}k}^d = P_{q\bar{q}}^d$ due to flavor independence of the kernel $P$. We have also re-labeled $P_{qq}^v/P_{qq}^s$ which indicates valence/sea in Eq.~(\ref{q-}) to 
$P_{qq}^c/P_{qq}^d$ to denote CI/DI.  

   It is clear from Eq.~(\ref{val}) that, with the proper definition of $q^v$ in Eq.~(\ref{lattice-valence}),
 there is no more valence flavor mixing as for $q^-$ in Eq.~(\ref{q-}). The evolution of the
 non-valance $q_i^{ds} - \bar{q}_i^{ds}$ in Eq.~(\ref{ds-diff}) has contributions from $q^{ds} - \bar{q}^{ds}$ and  the valence $q^v$ with different flavors. 
 Being $\mathcal{O}(\alpha_s^3)$, the kernel $P_{qq}^d - P_{q\bar{q}}^d$  is
small, but one needs to take into account the possibility of sizable intrinsic strange~\cite{Davidson:2001ji,Kretzer:2003wy} and charm    asymmetries~\cite{Brodsky:1980pb,Sufian:2020coz} which implies that $q^{ds} - \bar{q}^{ds}$ for
$q = u, d$ might be non-zero and larger in magnitude than those of the strange and charm.

\section{Are strange partons necessarily in the disconnected insertion?}   \label{strange}

\hspace{0.6cm}
  After the puzzles over the definition of the valence parton are resolved, there is a lingering question
as to why the strange and charm partons appear as the disconnected sea in the disconnected insertions only. Since the physical mass and matrix elements do not depend on the interpolation
field, as long as it has the right quantum numbers and a non-zero overlap with the hadronic state
under study, lattice practitioners usually adopt $uud$ interpolation field with $J^P = 1/2^+$ and
$I = 1/2$  for the proton for simplicity. In this case, the non-valence
strange and charm partons can only appear in DI. However, one could question if this is a special case due the restricted selection of the interpolation field. What if $\bar{s}s$ is included in the interpolation field in addition to $uud$? Would that entail strange to be a part of the CI and, perhaps,
a part of the valence?

To answer this question, we shall prove that the strange and charm will necessarily be in DI
with the special case of an interpolation field which includes the strange, i.e. 
$\mathcal{O} =uud\bar{s}s$. We shall first consider the two-point function
$C_{2pt} (\tau) = \langle O(\tau) O^{\dagger}(0)\rangle$.  With $\mathcal{O}(\tau) = e^{H\tau}O(0) e^{-H\tau}$, the 2-pt function can be written as
\begin{equation}
C_{2pt} (\tau) = \sum_n |\langle 0|O| n\rangle|^2 e^{-m_n\tau},
\end{equation}
after inserting a complete set of hadon states, such as $N, N\pi, N\pi\pi, NK\bar{K}, \Sigma K, R$, etc. with the nucleon quantum number. 
As shown in Fig.~\ref{5v-2pt}, the two-point function for $uud\bar{s}s$ has two path-integral diagrams
due to the two Grassmann number contractions. Fig.~\ref{5v-2pt} 
\begin{figure}[htbp] \label{5q-2pt}
\centering
\subfigure[]
{{\includegraphics[width=0.4\hsize]{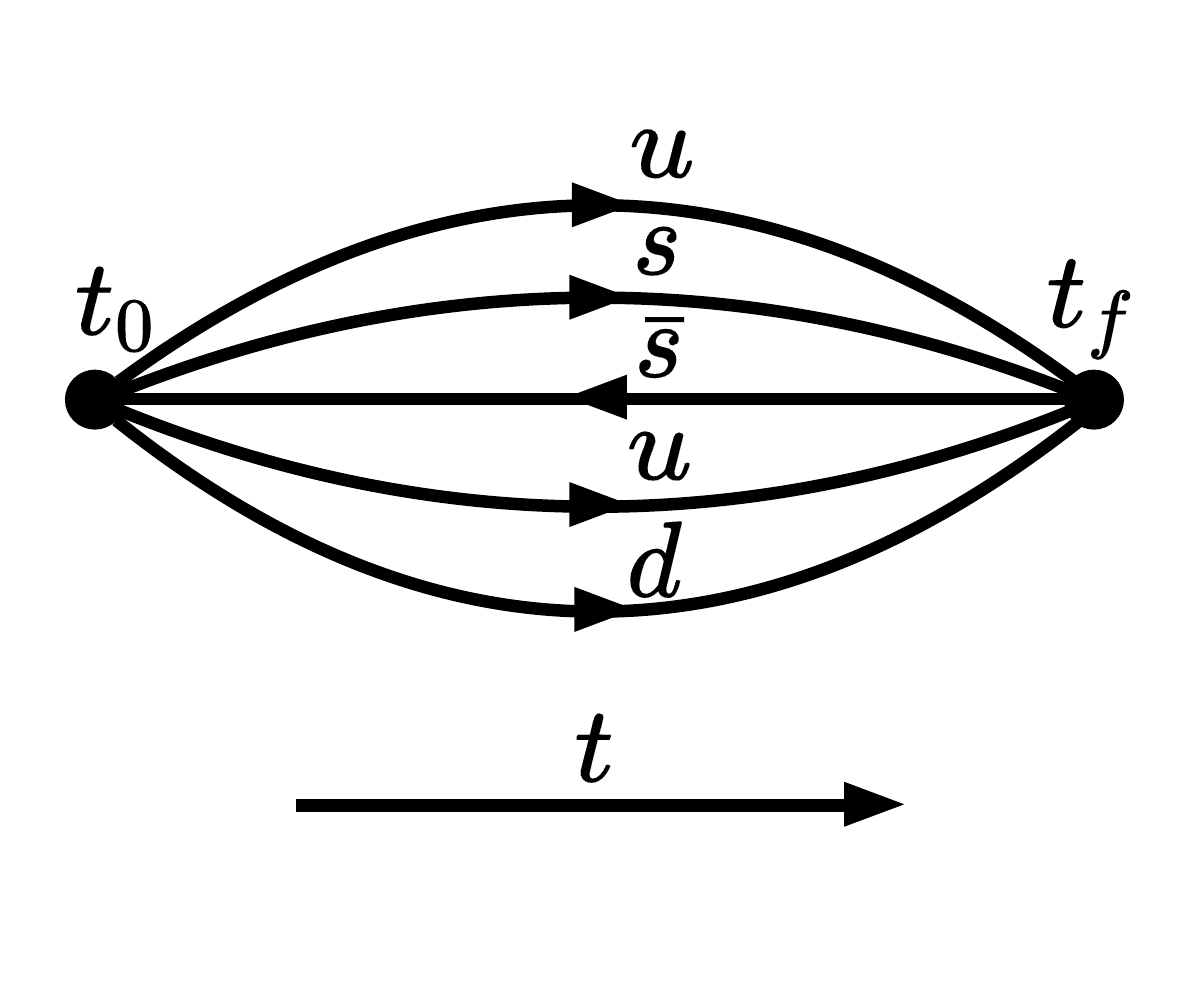}}
  \label{5v-2pt}}
\subfigure[]
{{\includegraphics[width=0.4\hsize]{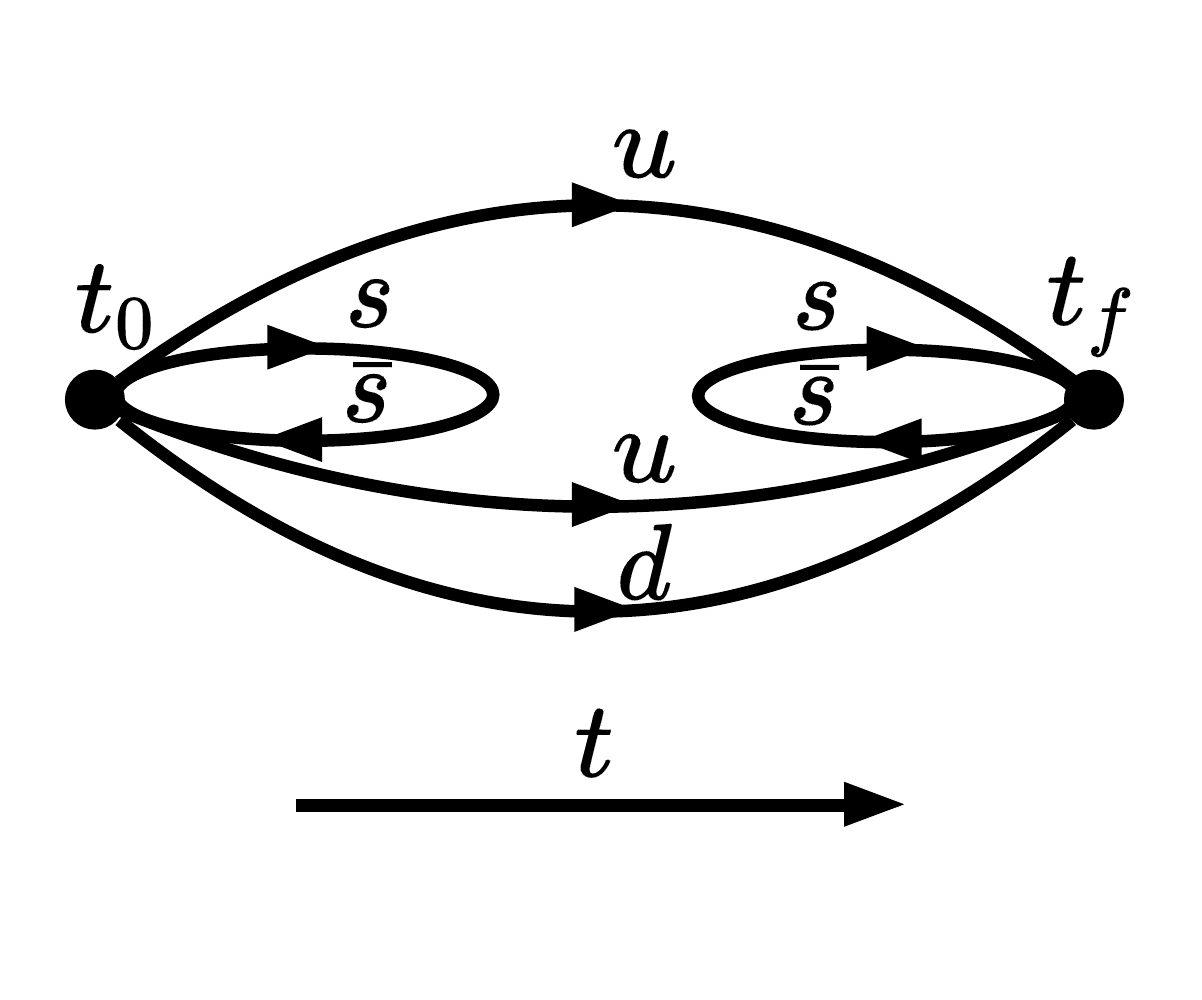}}
  \label{3v-2pt}}
%\vspace*{-0.6cm}
\caption{2-pt correlators for the $uud\bar{s}s$ interpolation field.  
 (a) for the case where all 5 valence quarks propagate from the source to the sink; (b) for 
 the case with $\bar{s}s$ annihilations.}
\end{figure}
in the left panel has 5 valence quarks at all times between the source and the
sink. When time separation $\tau = t_f - t_0$ is large, this part of the 2-pt correlator will have an asymptotic behavior
\begin{equation}
C_{2pt}^a (\tau)= W_{5q} e^{-M_{5q}\tau} + ...
\end{equation}
where  $...$ represents excited states. Having more valence quarks (one can count the minimum number of
quark propagators in a time slice) than that of the $uud$ interpolation field, $M_{5q}$, the lowest mass
of a state with a baryon and a meson(or mesons), will be higher than $M_N$, the nucleon mass. This is so, because the quark counting rule prescribes $\sim 300/500$ MeV constituent mass for each additional $q/s$ ($q = u,d$) 
quark~\footnote{The only exceptions we know are the scalar mesons 
below 1 GeV, which are believed to be the $Q^2\bar{Q}^2$ tetraquark 
mesoniums in the MIT bag model~\cite{Jaffe:1976ig} and the potential 
model~\cite{Liu:1981paa,Liu:1979en,Wong:1980xj}. They are verified in lattice calculations~\cite{Mathur:2006bs,Prelovsek:2010kg,Briceno:2016mjc,Guo:2018zss}.
These nonet mesoniums are below the $q\bar{q}$ nonet above 1 GeV~\cite{Mathur:2006bs,Cheng:2006hu,Liu:2007hma,Liu:2008ee,Draper:2008tp,Liu:2008bg,Buccella:2008zzb}.}.
This is verified with a lattice calculation with the $uud\bar{s}s$ interpolator, where the
$\bar{s}s$ is in the scalar channel~\cite{F.Lee}. This is also consistent with the recent lattice calculation~\cite{Yang:2020crz} which shows that the renormalized quark mass $m^R$ in the RI/MOM scheme with Landau gauge rises up to approach $\sim 300$ MeV for the scale below $\sim 500$ MeV. It coincides with the trace anomaly matrix element at different scales. Due to limited statistics, this is verified down to $\mu = 1.3$ GeV and has prompted the suggestion that trace anomaly and chiral symmetry breaking could be the origin of the constitute quark mass in the quark model. Therefore, one expects the mass of the 5-quark state to be higher than the nucleon, a 3-quark state, by about 600 MeV. 

The 2-pt function also has a insertion with $\bar{s}s$ annihilations in Fig.~\ref{3v-2pt}. Since the 
path-integral includes all possible paths, there are cases where the two $\bar{s}s$ loops do not overlap
(this happens when the $M_{5q}$ states from the source and sink are damped away exponentially),
leaving a gap in time which allows the 3-quark nucleon to emerge. In this the case, the lowest
mass state is the nucleon and the correlator for this part is
\begin{equation}
C_{2pt}^b (\tau)= W_N e^{-M_N \tau}+W_{5q}' e^{-M_{5q}' \tau} - W_{5q} e^{-M_{5q} \tau} + ...
\end{equation}
where $...$ includes the nucleon excited states, such as $\pi N$, Roper, etc. The sum is
then
\begin{equation}  \label{2pt}
C_{2pt}(\tau) = C_{2pt}^a (\tau) + C_{2pt}^b (\tau)= W_N e^{-M_N \tau}+W_{5q}' e^{-M_{5q}' \tau}  + ...
\end{equation}
where $M_{5q}'$ is the lowest full 5-quark baryon + meson state mass of the total $C_{2pt}(\tau)$ which is still an 
excited state of the nucleon. With $\Delta M = M_{5q}' - M_N > 0$, $C_{2pt}(\tau)$ is dominated by the nucleon state at large $\tau$.

The strangeness matrix element $\langle N|\bar{s}|\Gamma s|N\rangle$ is calculated from the ratio
of the 3-pt function to the 2-pt function. There are several contributions to the relevant 
3-pt function $C_{3p} (t_f - t, t - t_0)$ as illustrated in Fig.~\ref{5q-3pt}.
\begin{figure}[htbp]
\centering
\subfigure[]
{{\includegraphics[width=0.32\hsize]{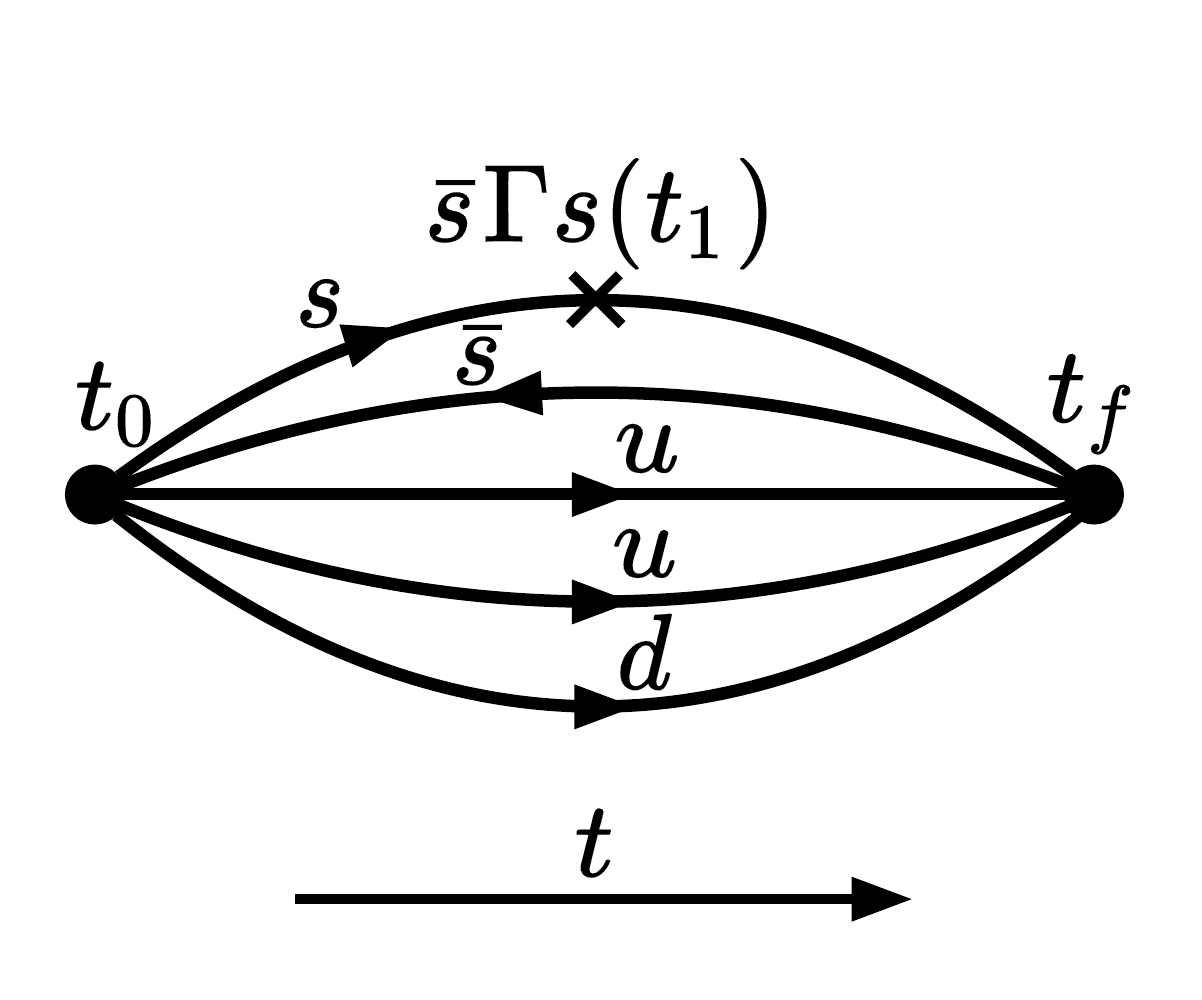}}
  \label{5v-3pt}}
\subfigure[]
{{\includegraphics[width=0.32\hsize]{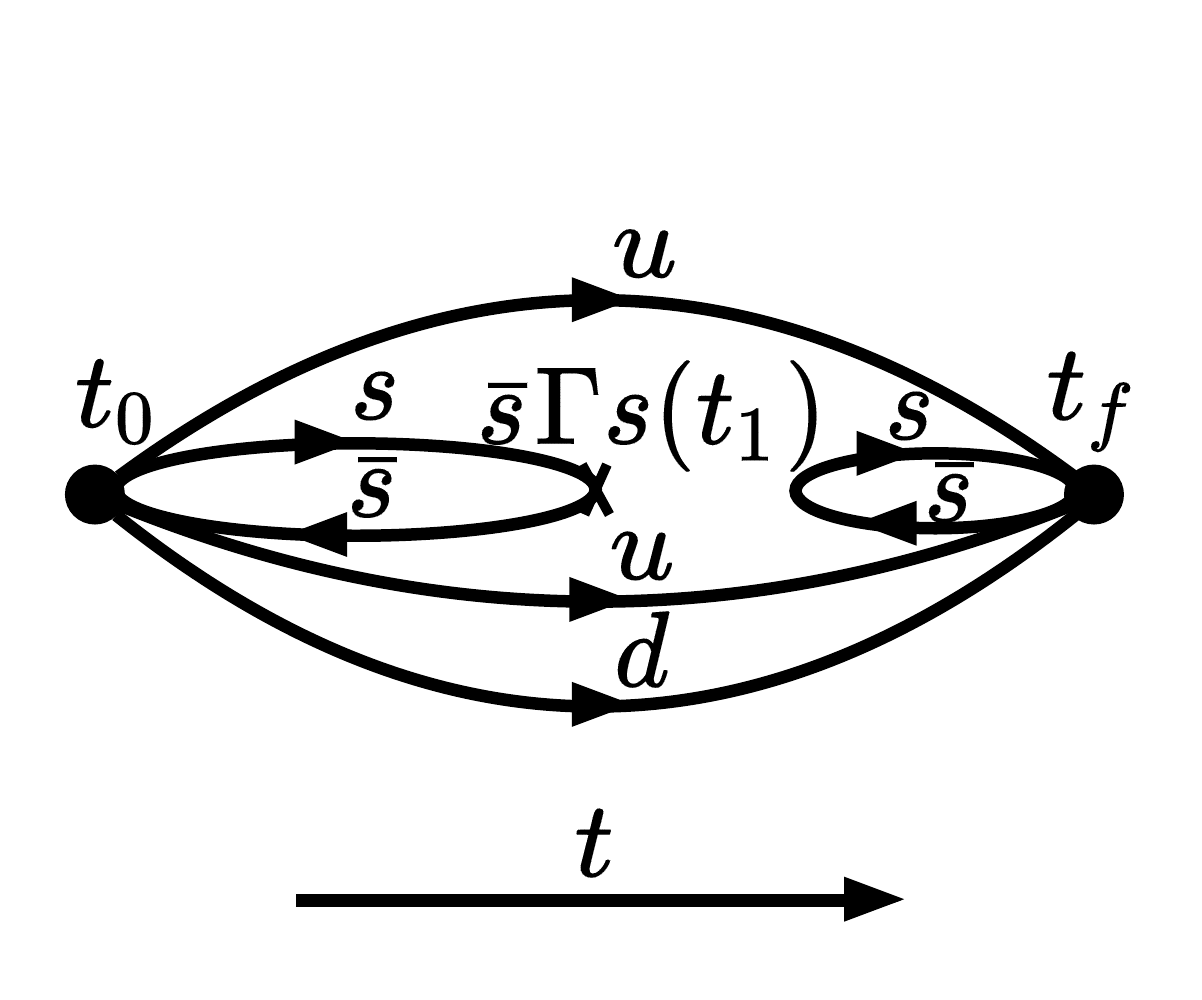}}
  \label{53v-3pt}}
\subfigure[]
{{\includegraphics[width=0.32\hsize]{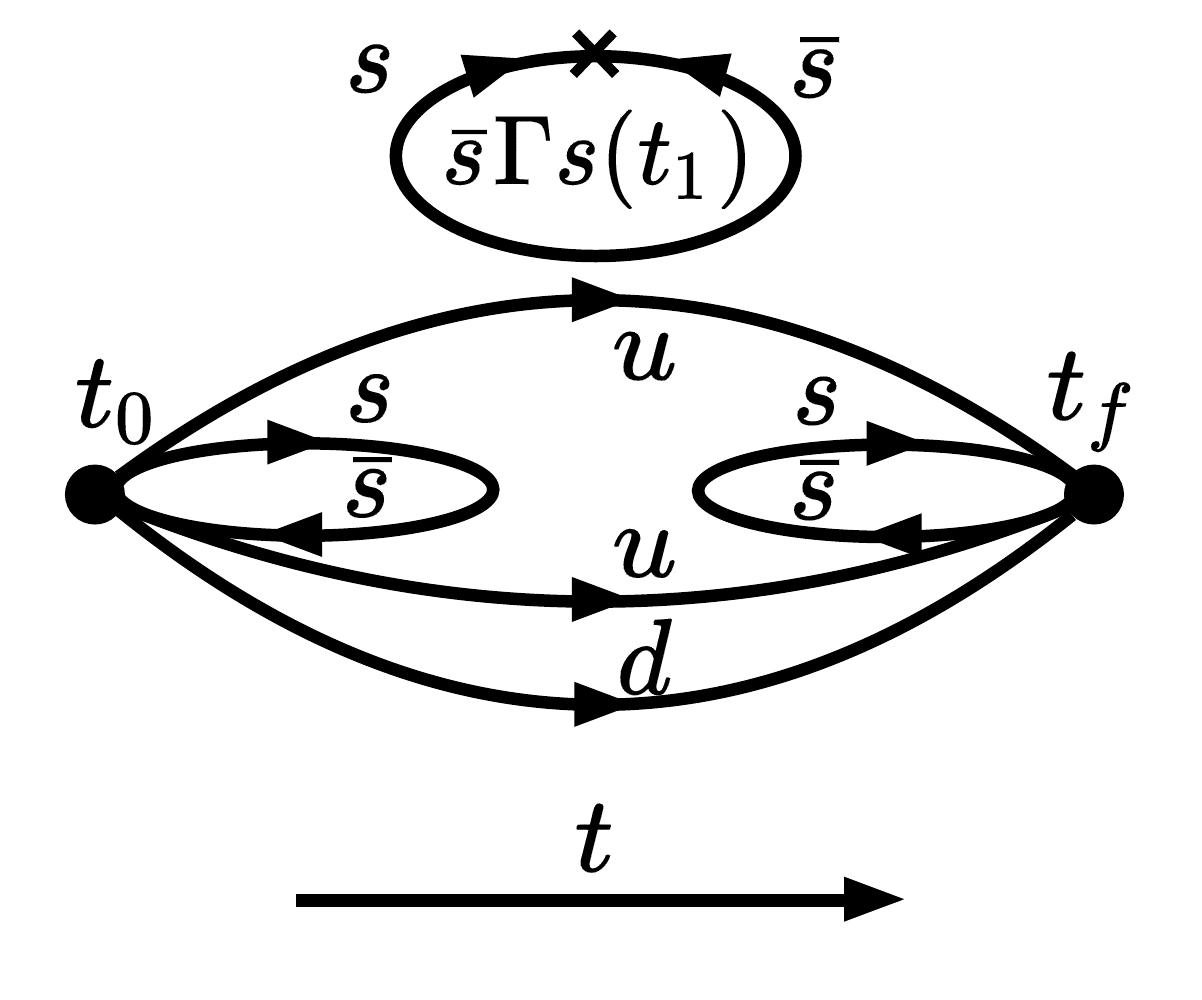}}
  \label{3v-3pt}}
%\vspace{-0.6cm}
\caption{Path-integral diagrams of 3-pt function with the $uud\bar{s}s$ interpolation field
and the insertion of the $\bar{s}\Gamma s$ current. 
 (a) is for the case where all 5 valence quarks propagate from the source to the sink; (b) is for 
 the case where the current is inserted on the annihilating $\bar{s}s$ pair from the sink or the source; and  (c) is for the case where the current forms a strange quark loop. 
 \label{5q-3pt}}
 \end{figure}
Inserting intermediate states between the source at $t_0$ and the operator $\bar{s}\Gamma s$ at $t$ 
and also between $t$ and the sink time $t_f$, the ratio is
\newpage
\begin{eqnarray}
\frac{C_{3pt} (t_f - t, t - t_0)}{C_{2pt}(t_f - t_0)} &=& \frac{W_{5q}}{W_N}\, \langle 5q|\bar{s}\Gamma s|5q\rangle\,
e^{- \Delta M \tau} + \sqrt{\frac{W_{5q}}{W_N}}\, (e^{-\Delta_M (t_f- t)} + e^{-\Delta_M (t - t_i)})\,
\langle 5q|\bar{s}\Gamma s|N\rangle \nonumber \\
&+& \langle N|\bar{s}\Gamma s|N\rangle+ ...,
\end{eqnarray}
for large time separations. The first term on the right is from Fig.~\ref{5v-3pt}, which has 5 valence quarks. Its ratio
involves the matrix element  $\langle 5q|\bar{s}\Gamma s|5q\rangle$ and an exponential factor 
$e^{- \Delta M \tau}$ which vanishes at large $\tau$. Similarly, the second term from
Fig.~\ref{53v-3pt}, where the current is inserted on the $\bar{s}s$ from the source or the sink, vanishes when $t_f - t$ and $t- t_i$ are large. The third term, where the $\bar{s}$ and $s$ in the current self contract to a loop, survives large time separation. It is the nucleon matrix element that we want. It still comes from the DI and is independent of the interpolation field.

It is straight-forward to generalize the proof for a general interpolation field $uud (\bar{q}q)^n f(G)$,
where $q$ is for any quark flavor, and $f(G)$ is a function of the gauge operators, as long as the
the interpolation field has the proton quantum number and has non-zero overlap with the
proton state.

\section{Summary}  \label{summary}
\hspace{0.6cm}
We show, in this work, that the problems posed by the phenomenological definition of the
valence parton as $q - \bar{q}$ in NNLO evolution is resolved by the appropriate valence definition
which involves only the parton and antipartons in the connected insertions from the QCD Euclidean
path-integral formulation of the hadronic tensor and the Feynman-$x$ approaches to PDFs. 
Since the path-integral formulation affords the separation of the connected and disconnected insertions,
a salient feature of the formulation, a total of 11 parton degrees of freedom are revealed, which are more than the ones (7 which are $u, \bar{u}, d, \bar{d}, s, \bar{s}$, and $g$) from the existing global analyses. These parton degrees of freedom  in the hadronic tensor, a Bjorken-$x$ approach, are shown to be the same as those from the Feynman-$x$ approaches via OPE (short distance Taylor expansion in the path-integral formalism). It is further proved that the non-valence strange and charm PDFs and their moments only contribute through the DI. 

It is encumbered upon global analyses to disentangle the connected sea from the disconnected
sea through the extended evolution equations~\cite{Liu:2017lpe}, so that lattice results can be compared to them for each degree of freedom. As for the moments, where precise lattice results are beginning to be available with all systematic errors taken into account, it is a good testing ground to make benchmark comparison between lattice calculations and experiments. With the CS and DS separated in global fits, the CI 
moments for the $u$ and $d$ and DI moments for $u,d$ and $s$ can be directly compared with lattice calculations, instead of being restricted to only the isovector and strange matrix elements.

\section{Acknowledgment}
The author is indebted to S. Brodsky, J.W. Chen, N. Christ, M. Diehl, T. Draper,Y. Hatta, X. Ji, F. X. Lee, J. Liang, P. Nadolsky, J.C. Peng, J.W. Qiu, A. Thomas, Y.B. Yang, and \mbox{C.P. Yuan} for insightful discussions. He also thanks Huey-Wen Lin and Martha Constantinou for allowing him to use their figures in Fig.~\ref{LatticeQPDF}.
 This work is partially support by the U.S. DOE grant DE-SC0013065 and DOE Grant No.\ DE-AC05-06OR23177 which is within the framework of the TMD Topical Collaboration.
This research used resources of the Oak Ridge Leadership Computing Facility at the Oak Ridge National Laboratory, which is supported by the Office of Science of the U.S. Department of Energy under Contract No.\ DE-AC05-00OR22725. This work used Stampede time under the Extreme Science and Engineering Discovery Environment (XSEDE), which is supported by National Science Foundation Grant No. ACI-1053575.
We also thank the National Energy Research Scientific Computing Center (NERSC) for providing HPC resources that have contributed to the research results reported within this paper.
We acknowledge the facilities of the USQCD Collaboration used for this research in part, which are funded by the Office of Science of the U.S. Department of Energy.

%-----------------------------------------------------------------------

\end{document}